\documentclass[final]{siamart1116}
\usepackage{etex}
\usepackage[spanish,english]{babel}
\usepackage{amsfonts} 
\usepackage{amsmath}
\usepackage{amssymb}
\usepackage{booktabs}
\usepackage{epstopdf}
\usepackage{graphicx}
\graphicspath{{./Figures/}}
\usepackage{subfigure}
\usepackage{colortbl}
\usepackage{pst-all}
\usepackage{wasysym}
\usepackage{mathrsfs}
\usepackage{float}
\usepackage{verbatim}
\usepackage{upgreek}
\usepackage{todonotes}
\usepackage{color}
\usepackage[applemac]{inputenc}

\usepackage{array}
\newcolumntype{L}[1]{>{\raggedright\let\newline\\\arraybackslash\hspace{0pt}}m{#1}}
\newcolumntype{C}[1]{>{\centering\let\newline\\\arraybackslash\hspace{0pt}}m{#1}}
\newcolumntype{R}[1]{>{\raggedleft\let\newline\\\arraybackslash\hspace{0pt}}m{#1}}

\usepackage{tikz}
\usetikzlibrary{arrows,shapes,backgrounds}

\newcommand{\pgftextcircled}[1]{                                                                    
    \setbox0=\hbox{#1}%
    \dimen0\wd0%
    \divide\dimen0 by 2%
    \begin{tikzpicture}[baseline=(a.base)]%
        \useasboundingbox (-\the\dimen0,0pt) rectangle (\the\dimen0,1pt);
        \node[circle,draw,outer sep=0ex,inner sep=0.1ex] (a) {#1};
    \end{tikzpicture}
}

\let\oldsqrt\sqrt
\def\sqrt{\mathpalette\DHLhksqrt}
\def\DHLhksqrt#1#2{%
\setbox0=\hbox{$#1\oldsqrt{#2\,}$}\dimen0=\ht0
\advance\dimen0-0.2\ht0
\setbox2=\hbox{\vrule height\ht0 depth -\dimen0}%
{\box0\lower0.4pt\box2}}

\newcommand{\bsub}{\begin{subequations}}
\newcommand{\esub}{\end{subequations}$\!$}
\renewcommand{\theequation}{\arabic{section}.\arabic{equation}}
\renewcommand{\thesection}{\arabic{section}}

\newcommand{\eps}{\varepsilon}

\DeclareMathOperator{\sgn}{sgn}
\DeclareMathOperator{\var}{var}

\graphicspath{{Figures/}}

\newcommand{\TheTitle}{Ducks in space: from nonlinear absolute instability to
noise-sustained structures in a pattern-forming system}

\newcommand{\TheAuthors}{D. Avitabile, M. Desroches, E. Knobloch and M. Krupa}
\headers{Ducks in space}{\TheAuthors}

\title{{\TheTitle}}

\author{
  Daniele Avitabile
  \thanks{Centre for Mathematical Medicine and Biology, School of
  Mathematical Sciences, University of Nottingham, University Park, Nottingham, NG7
  2RD, 
    \email{daniele.avitabile@nottingham.ac.uk},
    ORCID ID \url{http://orcid.org/0000-0003-3714-7973}.
  }
  \and
  Mathieu Desroches
  \thanks{MathNeuro Team, Inria Sophia Antipolis M\'{e}diterran\'{e}e, 2004 Route des
  Lucioles, BP93, 06902 Valbonne cedex, France
    ORCID ID \url{http://orcid.org/0000-0002-9325-4207}.
  }
  \and
  Edgar Knobloch
  \thanks{Department of Physics, University of California, Berkeley, CA 94720, USA.
  }
  \and
  Martin Krupa
  \thanks{Universit\'e C\^ote d’Azur \& MathNeuro Team, Inria Sophia Antipolis M\'{e}diterran\'{e}e, 2004 Route des
  Lucioles, BP93, 06902 Valbonne cedex, France
    ORCID ID \url{http://orcid.org/0000-0001-9330-4705}.
  }
}

\ifpdf
\hypersetup{pdftitle={\TheTitle},pdfauthor={\TheAuthors}}
\fi

\begin{document}
\maketitle

  
\begin{abstract}
  A subcritical pattern-forming system with nonlinear advection in a bounded domain
  is recast as a slow-fast system in space and studied using a combination of
  geometric singular perturbation theory and numerical continuation. Two types of
  solutions describing the possible location of stationary fronts are identified,
  whose origin is traced to the onset of convective and absolute instability when the
  system is unbounded.  The former are present only for nonzero upstream boundary
  conditions and provide a quantitative understanding of noise-sustained structures
  in systems of this type.  The latter correspond to the onset of a global mode and
  are present even with zero upstream boundary condition. The role of canard
  trajectories in the nonlinear transition between these states is clarified and the
  stability properties of the resulting spatial structures are determined. Front
  location in the convective regime is highly sensitive to the upstream boundary
  condition and its dependence on this boundary condition is studied using a
  combination of numerical continuation and Monte Carlo simulations of the partial
  differential equation. Statistical properties of the system subjected to random or
  stochastic boundary conditions at the inlet are interpreted using the deterministic
  slow-fast spatial-dynamical system.
\end{abstract}

\section{Introduction}
Systems of hydrodynamic type with through-flow are well-known to be highly sensitive to
the upstream boundary condition. In some systems this is a consequence of the non-normality
of the linear stability operator; such operators can greatly amplify small perturbations even
when the base flow is linearly stable, resulting in transient amplification \cite{ReddyTrefethen}. 
In others it is a consequence of convective instability within the linear stability problem, 
i.e., an instability that grows in an appropriately moving reference frame, although it decays 
at any fixed downstream location. As a result, continued injection of fluctuations at the
upstream boundary leads to the presence of a {\it noise-sustained} structure downstream \cite{Deissler87}.
When the source of fluctuations is shut off these structures are swept out of the system,
i.e., the downstream disturbance at any fixed location decays to zero. 
In many systems both processes may be active simultaneously. 

When perturbations grow at fixed location we speak
instead of absolute instability. Evidently, convective instability must precede absolute
instability since the latter requires a sufficiently large growth rate that an initially
localized disturbance can expand upstream against downstream advection. Structures associated
with absolute instability are no longer sensitive to the upstream boundary condition. 

The present paper seeks to investigate the consequences of convective instability in nonlinear
systems. Of particular interest are systems exhibiting subcritical absolute instability, a
situation that arises frequently in shear flow problems \cite{Tuckerman14}. In such
systems the base state may be convectively unstable but finite amplitude
perturbations can result in nonlinear absolute instability, i.e., nonlinear states
that are no longer swept downstream. We are particularly interested in elucidating
the transition from convective to absolute instability in the nonlinear regime, and
the associated changes in the sensitivity of the observed structures to the upstream
boundary condition.

Nonlinear noise-sustained structures have been studied by direct numerical simulations
while most of the large body of work associated with shear flows focuses on structures
generated by nonlinear absolute instability. However, the transition between the two
has not, to our knowledge, been the subject of any past study. Owing to the difficulty
of the subject we choose to investigate here a particular system that is known to
display the required properties. For this purpose we select a model first studied by
Chomaz \& Couairon \cite{Chomaz97,Chomaz99}:
\begin{equation}\label{CCmodel}
  A_t=A_{xx}-(U-\alpha A^2)A_x+\mu A-A^3.
\end{equation}
In this partial differential equation (PDE) $A$ is a real variable representing the 
amplitude of a nonlinear state and $U$, $\alpha$ and $\mu$ are control parameters. The parameter $U$ measures the
strength of advection while $\mu$ is considered to be the principal bifurcation parameter:
on the real line the onset of convective instability of the trivial state $A=0$ occurs
at $\mu=0$ while the onset of absolute instability takes place at $\mu=\mu_{\rm a}:=U^2/4$.
Finally, the coefficient $\alpha$ quantifies the importance of the nonlinear correction
to the advection velocity $U$. Although this model cannot be derived from the Navier-Stokes
equation by standard multiple scale techniques (except near degeneracies) it is faithful
to the physics of this class of problems as well as being simple enough to be amenable to
relatively complete analysis, and in particular to phase plane techniques.

Chomaz \& Couairon posed the above problem on a finite domain, $x \in [0,L]$, subject
to Dirichlet boundary conditions $A(0,t) = A(L,t) = 0$, and studied the strong
advection regime $U \gg 1$. The trivial homogeneous steady state is unstable for
sufficiently large values of the control parameter $\mu$, and a subcritical bifurcation
at $\mu=\mu_{\rm g}>\mu_{\rm a}$ gives rise to a branch of stationary states with sharp spatial
gradients and an $\mathcal{O}(\sqrt{\mu})$ amplitude
(see Figure~\ref{fig:chomazResults}(a), which is adapted from a sketch by Chomaz
\& Couairon~\cite{Chomaz99}).
We study this model in the case where the inlet boundary condition is small, but
nonzero, $0 < |A(0,t)| = \eta \ll 1$, while also focusing on the strong advection regime
$U \gg 1$. For numerical illustrations we take $U=12$ (hence $\mu_\textrm{a} = 36$)
  and $L=10\pi$, unless otherwise stated. The remaining parameters are assumed to be of order
one. The main results of our paper can be summarized as follows:
\begin{figure}[!h]
  \centering
  \includegraphics[width=8cm]{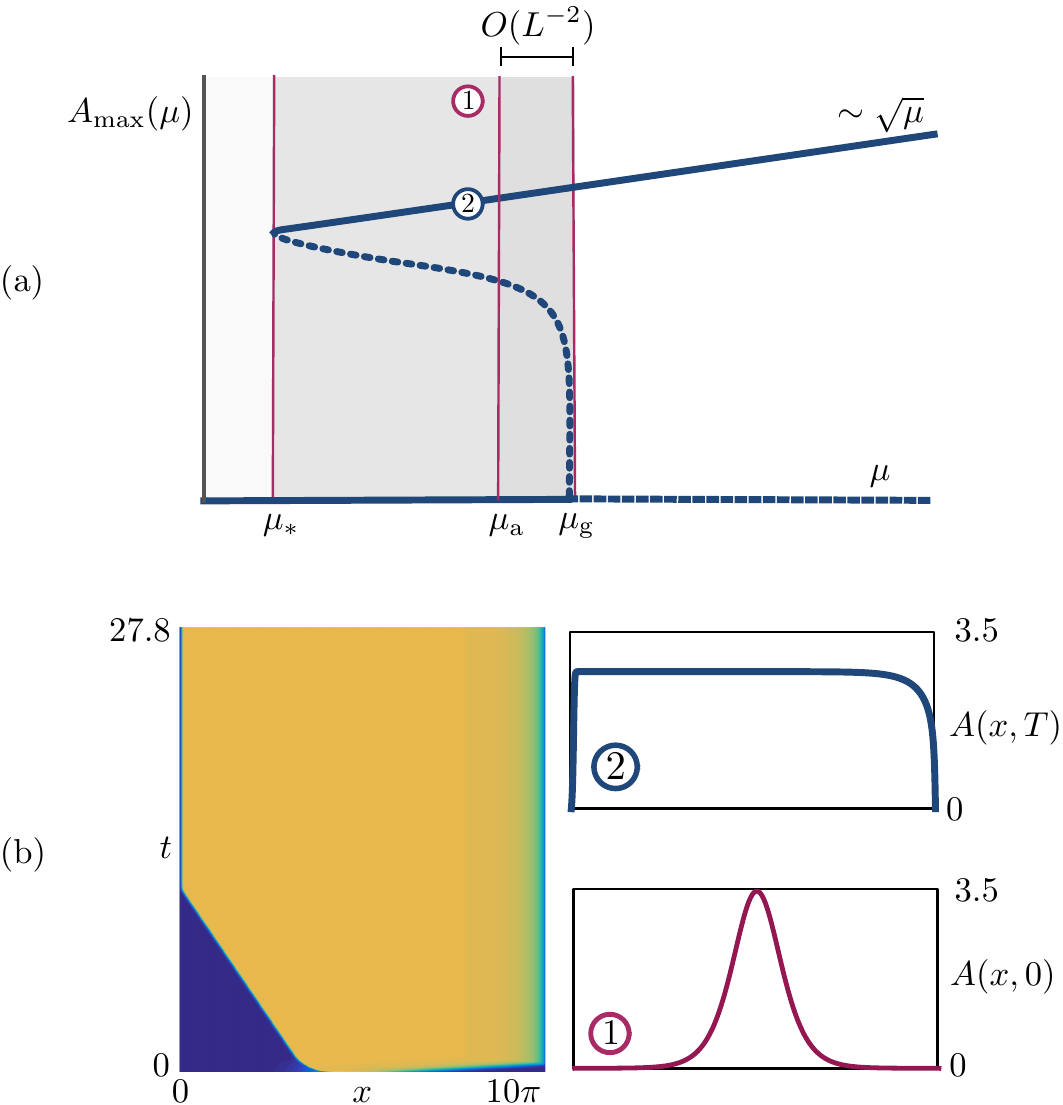}
  \caption{
    (a) Sketch of the bifurcation diagram of steady states of Eq.~\eqref{CCmodel} with
    Dirichlet boundary conditions  $A(0,t) = A(L,t)\equiv 0$, adapted
  from~\cite{Chomaz99}.
    (b) Time simulation of~\eqref{CCmodel} with Dirichlet boundary conditions. Left: space-time
    plot for $(x,t)\in [0,L] \times [0,T]$ with $L=10 \pi$ and $T=0.21$. Right: initial
    (1) and final (2) profiles of the time simulation. These states are also marked on
    the bifurcation diagram in (a). Parameters: $\mu = 7.4<\mu_\mathrm{a}$, $\alpha = 5$,
    $U=12$. 
    }\label{fig:chomazResults}
\end{figure}

\begin{figure}[!h]
  \centering
  \includegraphics{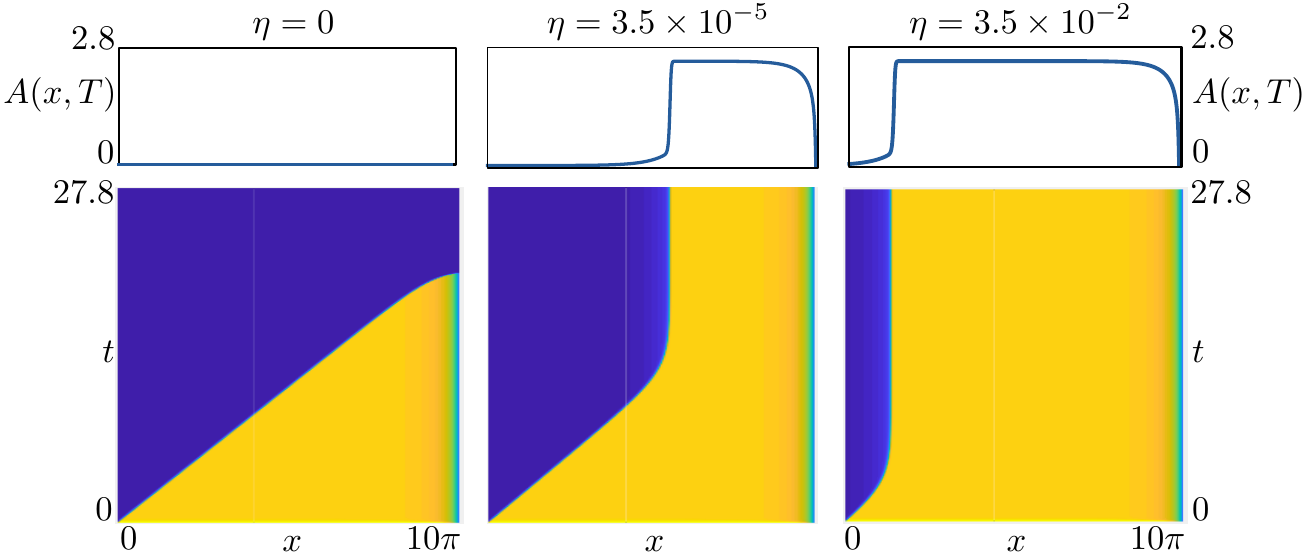}
  \caption{Small variations of the inlet boundary condition $A(0,t) = \eta$ have a
    dramatic impact on the steady states attained in Eq.~\eqref{CCmodel}.
    Left: when we time-step~\eqref{CCmodel} starting from a solution with zero inlet
    boundary condition, $\eta=0$, the system evolves towards the trivial steady state.
    Center: with $\eta=3.5 \times 10^{-5}$ a new solution is selected. Right: a similar
    final steady state, with a longer plateau, is attained with $\eta=3.5 \times 10^{-2}$.
    Parameters: $\mu = 6$, $\alpha = 5$, $U=12$, with $(x,t) \in [0,10\pi] \times
    [0,2.5]$. In all cases $A(L,t)=0$.}
  \label{fig:timeStepA0Compound}
\end{figure}

\begin{figure}[!h]
    \centering
    \includegraphics{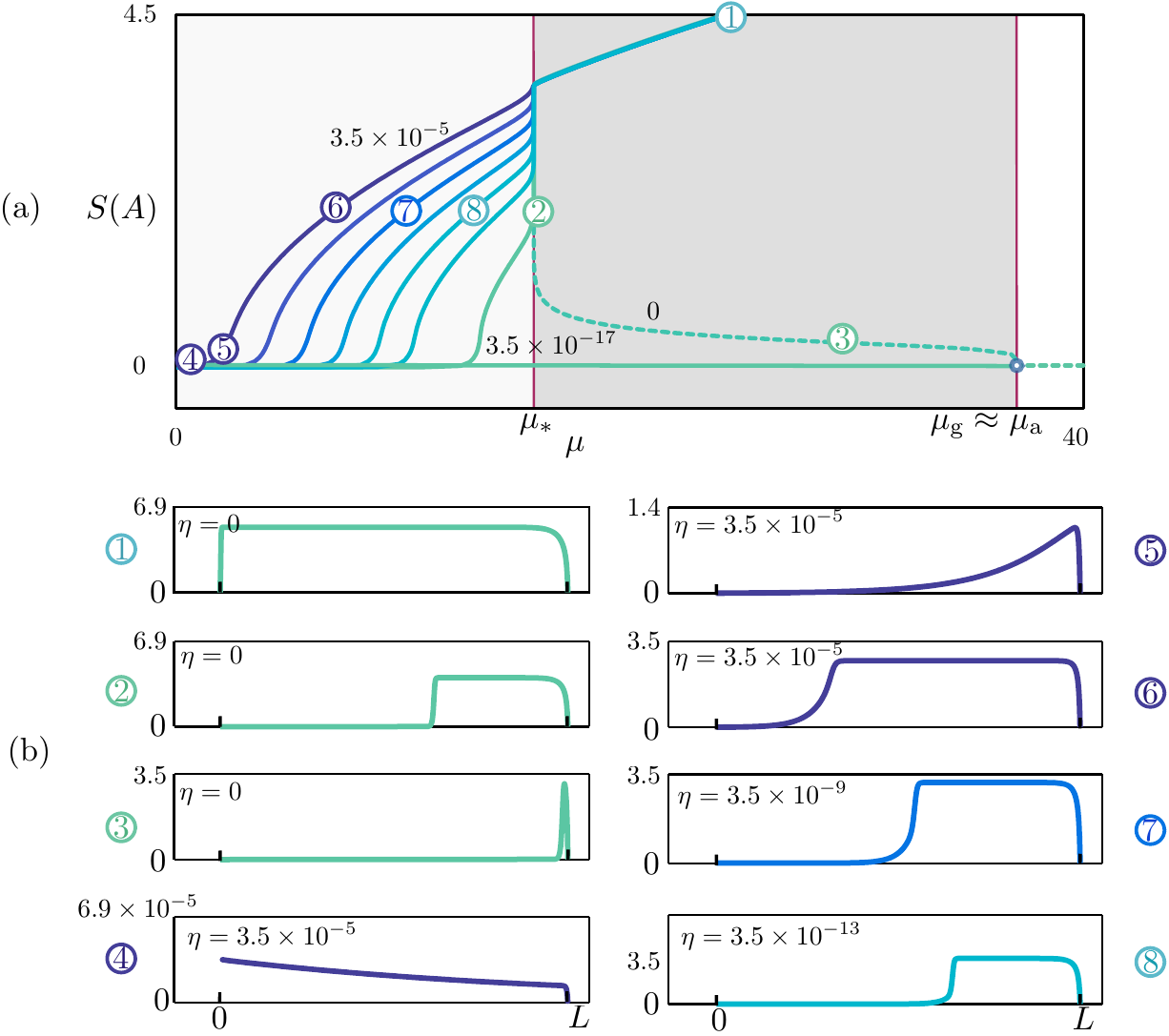}
    \caption{
    (a) Branches of steady states of Eq.~\eqref{CCmodel} for $\alpha = 2$, $U =
    12$, $L = 10 \pi$, represented using the solution measure $S(A) \equiv -\sgn A'(L) \Vert A \Vert_2$
    along the vertical axis, for various inlet boundary conditions $A(0,t) = \eta \in
    [0,3.5 \times 10^{-5}]$ and $A(L,t)=0$.
    Here the prime denotes differentiation with respect to $x$ and $\Vert \cdot \Vert_2$
    is the standard $L_2$ norm. Stable (unstable) branches are indicated via solid
    (dashed) lines. (b) The solution profiles $A(x)$, $x \in [0,L]$, corresponding to the
    locations indicated in panel (a). The vertical scale is indicated for each profile separately.
    The states in $\mu < \mu_\ast$ originate in the convective instability of $A=0$,
    while the states in $\mu > \mu_\ast$ are the consequence of absolute instability.
    }\label{mubrPDEAlt}
\end{figure}

\maketitle
\begin{enumerate}
  \item The model~\eqref{CCmodel} displays high sensitivity to $\eta$, as documented in
    Figure~\ref{fig:timeStepA0Compound}, where the formation of a front between the
    base state and the finite amplitude state is shown in a sequence of space-time
    plots for different values of $\eta$.

  \item Numerical continuation of steady states for $\eta \neq 0$ reveals the presence
    of two types of finite amplitude states, one originating in the convective
    instability threshold $\mu=0$ and the other in the absolute instability threshold at
    $\mu=\mu_{\rm g}$. The former reveal extreme sensitivity to the value of $\eta$
    and disappear abruptly as $\eta\rightarrow 0^+$ (Figure~\ref{mubrPDEAlt}).

  \item The observed sensitivity to the parameter $\eta$ finds a natural explanation
    in the presence of canard segments (see below for a definition) on
    solutions of the spatial boundary value problem (BVP) for the steady states of~\eqref{CCmodel}.
    These results are obtained by recasting the BVP as a slow-fast system in space subject to the
    boundary conditions $A(0,t) = \eta$, $A(L,t) = 0$. The solutions of this BVP
    correspond to finite length trajectories of a slow-fast spatial-dynamical system
    of van der Pol type. 

   \item The stability in time of all steady solutions of the BVP is determined and
     corroborated using direct numerical simulations.
      
   \item Numerical evidence is presented that demonstrates that the statistics of the
     front location in the convectively unstable regime can be explained solely in terms
     of the properties of the new branches of steady states in the convectively unstable
     regime, present for $\eta \neq 0$.
\end{enumerate}

As emphasized above, the slow-fast structure of the van der Pol spatial-dynamical
system plays an important role in understanding the dynamics of the system. In
recent years there has been a great deal of interest in slow-fast systems 
\cite{Desroches12}. Among the many phenomena such systems exhibit is the canard
phenomenon~\cite{Benoitetal81,KrupaSzmolyan01}. This term is used to refer to
the abrupt growth of a limit cycle that occurs, in planar systems, in exponentially
narrow parameter intervals. Within this interval the limit cycle follows for a time 
a repelling slow manifold~\cite{Fenichel79}. Such cycles are known to occur in the van der Pol
system, and have been studied intensively in the past~\cite{Benoitetal81,KrupaSzmolyan01}.
In the following we use the term {\it canard segment} to refer to the part of a
solution that follows a repelling slow manifold; depending on the situation, such a
segment can be part of a periodic solution or not. A {\it maximal canard segment}
follows a repelling slow manifold for as long as it exists; for example, in the van
der Pol oscillator, a maximal canard segment (which is part of a limit cycle) follows
the middle (repelling) branch of the cubic nullcline from the lower fold point all
the way to the upper one.
Phenomena of this type are invariably studied in the time domain, however, and in this paper 
we seek to extend the slow-fast decomposition to solutions that evolve in space, forming, 
for example, a front separating two distinct states of the system. However, in the
spatial-dynamics formulation one is typically not interested in a periodic state: 
such systems are typically BVPs with non-periodic boundary conditions.
For example, orbits corresponding to steady states of~\eqref{CCmodel} may be required
to satisfy $A(0) = \eta \neq 0 = A(L)$ and the familiar van der Pol canard cycles are 
not observable in this setting. We eschew here the rigorous construction of such orbits
and instead employ phase plane analysis combined with numerical bifurcation analysis to 
construct relevant orbits of the slow-fast system and to provide numerical evidence for 
the existence of canard segments.

In addition, we study the stability \textit{in time} of the steady states of~\eqref{CCmodel};
in other words, we are interested in canard-like phenomena in the context of PDEs. Amongst the
few studies in this direction, we mention \textit{canard traveling waves}, which have been
analyzed in a moving frame as homoclinic connections with a canard segment
\cite{buric06,Harterich03,SSS03}, shock-like structures in PDEs which can also be
interpreted in terms of canards~\cite{Carter15,WP} and canard-related bifurcation
delay in reaction-diffusion PDEs \cite{DPK,KathMurray85}. In recent work we
classified for the first time spatio-temporal canards in an
infinite-dimensional dynamical system arising in neuroscience applications, using
an interfacial dynamics approach~\cite{Avitabile17}. We leave the mathematical study of purely temporal
canards in PDEs, such as those reported by Gandhi et al.~\cite{Gandhi16}, to
future work.

The paper is organized as follows. In Section \ref{CC} we provide a brief summary
of the results of Chomaz \& Couairon. In Section \ref{scaling} we show how the equation can
be cast into a slow-fast system suitable for phase-plane analysis. The phase-plane analysis
itself is presented in Section \ref{phaseplane} with additional details relegated to Appendix
\ref{app:pplane}. Next, we discuss the stability properties of these states in time, and then
perform direct numerical simulations of Eq.~\eqref{CCmodel} for both time-periodic and stochastic
inlet boundary conditions (Section \ref{sec:simulations}) and thereby establish a relation between
the stationary convective instability states shown in Figure \ref{mubrPDEAlt}(a) and the
noise-sustained structures observed in this system (Figure \ref{fig:timeStepA0Compound}). 
Throughout the paper we focus on revealing the origin of the extreme sensitivity of this
model problem to the inlet boundary conditions in the convectively unstable regime (Figure \ref{fig:timeStepA0Compound}).

\section{Basic properties of the model}\label{CC}

In this Section we summarize the basic properties of the model \eqref{CCmodel} on large
but finite domains. As already explained the bifurcation parameter $\mu$ controls
the stability of the trivial state $A = 0$: on the real line, homogeneous solutions
grow when $\mu>0$ and saturate at amplitude $\pm\sqrt{\mu}$ owing to the nonlinear
term $-A^3$. When $\alpha=0$ the advection term $UA_x$ represents advection of spatial
inhomogeneities to the right at speed $U$. As
a result near $0 < \mu \ll 1$ infinitesimal spatially localized disturbances of
$A=0$ travel to the right more rapidly than they grow and at each fixed location the
disturbance decays to zero as $t\rightarrow\infty$. Thus, despite the growth of the
instability in an appropriately moving frame, the state $A = 0$ remains stable in this
sense. In the fluid mechanical literature the point $\mu=0$ is referred to as the
threshold for convective instability. The term $\alpha>0$ does not change this linear
picture but it does imply that the advection speed is reduced as the amplitude of the
state grows. Thus nonlinear states are not necessarily advected downstream and can in
fact invade the whole domain (Figure~\ref{fig:chomazResults}(b)). In the following we
refer to this behavior as nonlinear absolute instability by analogy with the notion of
(linear) absolute instability which arises at $\mu=\mu_{\mathrm{a}} := U^2/4$. At this
parameter value the growth rate of a localized infinitesimal perturbation of $A=0$ is
able to compete for the first time with the rate at which it is swept downstream, i.e.,
at $\mu=\mu_{\mathrm{a}}$ the upstream spreading speed exactly balances the downstream advection.

In systems of this kind, boundaries play a crucial role: linear theory shows that in
the presence of (nonperiodic) boundary conditions that preserve the base state $A=0$,
this state loses stability at $\mu_\mathrm{g}$, corresponding to the presence of a
neutrally stable global eigenmode $A_\mathrm{g}(x)$. In domains of large length $L$
one finds that $\mu_\mathrm{g}=\mu_{\mathrm{a}}+O(L^{-2})$ \cite{Chomaz99,Tobias98}.
In such domains the threshold of linear instability is thus delayed, by an $O(1)$
amount, from the threshold $\mu=0$ for the onset of instability in the corresponding
problem with periodic boundary conditions. Thus for $\mu<\mu_\mathrm{g}$
infinitesimal perturbations of $A=0$ decay, albeit on an $O(L)$ time scale, but
for $\mu>\mu_\mathrm{g}$ they grow. For $\mu=\mu_\mathrm{g}+O(L^{-5})$ the solutions of
the nonlinear problem resemble the linear theory eigenmode $A_\mathrm{g}(x)$ at
$\mu=\mu_\mathrm{g}$~\cite{Chomaz99,Tobias98}; this eigenmode is typically compressed
against the downstream wall and we call it a wall mode. As $\mu$ increases further,
an abrupt transition takes place whereby the domain fills with the $A\ne0$ state,
except for a residual defect near the upstream boundary \cite{Tobias98}.

The above scenario appears to be typical of systems with broken reflection
symmetry, confined in a finite domain, provided the primary bifurcation is
supercritical. However, as already mentioned, the system (\ref{CCmodel}) is
expected to become strongly subcritical when $\alpha$ is sufficiently large (in
fact, $\alpha>6/U$ \cite{Chomaz99}) and this fact suggests that a different
domain-filling scenario may take place. Figure \ref{fig:chomazResults}(a) sketches
the typical bifurcation diagram one obtains for $\alpha>6/U$. The figure shows the
maximum value of $A$ as a function of $\mu$ and reveals that when $A(0,t)=0$ the
branch of global solutions bifurcates subcritically from $\mu=\mu_{\rm g}$ before turning
around towards larger values of $\mu$ at a fold located at $\mu=\mu_\ast$. For $\mu<\mu_\ast$
no steady, i.e., persistent, states of the system exist and all initial perturbations
ultimately decay to $A=0$. However, as shown below, as soon as $ 0 < |A(0,t)| \ll 1$,
this is no longer the case and a branch of steady states exists throughout $0<\mu<\mu_\ast$.

A remarkable fact is that these branches are extremely sensitive to the exact value
of the inlet boundary condition $A(0,t)$. We illustrate this fact by
explicit computations of steady states of Eq.~\eqref{CCmodel} with boundary
conditions $A(0,t)=\eta$, $A(L,t)=0$, for moderate values of the advection speed
$U$ and $\alpha=2$ (Figure \ref{mubrPDEAlt}). The figure reveals the presence of nonzero steady
solutions in $\mu<\mu_*$ even for extremely small values of $\eta$
($3.5 \times 10^{-17}\le\eta\le 3.5 \times 10^{-5}$),
together with sample solution profiles at locations indicated in the figure. Unlike
Figure \ref{fig:chomazResults}(a), Figure \ref{mubrPDEAlt}(a) shows the quantity
$S(A) \equiv -\sgn A'(L) \Vert A \Vert_2$ as a function of $\mu$, where the prime
denotes differentiation with respect to $x$ and $\Vert \cdot \Vert_2$ is the standard
$L_2$ norm, and does so for various inlet boundary conditions $\eta \in [0,3.5 \times 10^{-5}]$.

The new solution measure unfolds the neighbourhood of $\mu_\ast$ for $\eta=0$,
revealing a steep vertical branch, along which the position of the front varies
continuously (Figure~\ref{mubrPDEAlt}(b)). The figure shows that the fold at
$\mu=\mu_\ast$ on the $\eta=0$ solution branch is in fact highly degenerate and
that all the $\eta\ne 0$ branches accumulate on it. It turns out (see below) that
the degeneracy is a reflection of an orbit flip in the spatial dynamics description of
the system \cite{Chomaz97}, while the existence of the finite amplitude states for
$0<\mu<\mu_\ast$ is related to canard-like behavior present in this description. We
also provide a physical interpretation of the states found for $\mu<\mu_\ast$, and
relate their properties to the noise-sustained structures present in convectively
unstable systems \cite{Deissler87}.

Both boundary layers and interior layers occurring in singularly perturbed
nonlinear BVPs have in fact been studied extensively over many
years~\cite{Chang2012,DeJager1996a}. In addition, the work by Gorelov and
Sobolev~\cite{Gorelov91,Gorelov92} analyzes canard solutions to BVPs in ODE models of
combustion. A
recent review~\cite{omalley}
provides an excellent guide to the literature on BVPs with Dirichlet boundary
conditions and the connection between such systems and the canard phenomenon,
albeit almost exclusively for linear systems. In the present paper we extend
this type of discussion not only to strongly nonlinear situations, but also
explain how the bifurcation structure associated with canard-type phase plane
trajectories affects the {\it temporal} dynamics of the PDE, subject to variable
inlet boundary conditions. The spatial orbits containing canard segments considered
here are in fact transient solutions to the spatial dynamics problem, and so are
unrelated to the classical periodic canard cycles of the (forced) van der Pol oscillator,
which are asymptotic states.

\section{The slow-fast system}\label{scaling}

In this paper we seek to extend the above results in two directions, to determine the nonlinear
states (if any) present in the convectively unstable regime, and to relate these states to the
states generated by the subcritical absolute instability at $\mu=\mu_{\rm g}$. For this purpose
we consider Eq.~\eqref{CCmodel} in the limit of large advection and introduce the small parameter
\[
0 < \eps = U^{-2} \ll 1.
\]
To retain a significant nonlinear contribution to the advection speed we scale the amplitude $A$
as follows, $A = \eps^{-1/4} A'$, where by assumption $A'=O(1)$, i.e., we focus on appropriately
large amplitudes $A$. The diffusion term is retained at leading order provided $x = \eps^{1/2} x'$,
where $x'=O(1)$, i.e., we focus on small spatial scales $x$, of order $\eps^{1/2}$, and hence consider
domains of size $L=\eps^{1/2}\ell$, where $\ell$ is formally $O(1)$ but is large compared to the scale
on which the front evolves. Time evolution then takes place on an $O(\eps)$ time
scale (see Figures~\ref{fig:chomazResults} and \ref{fig:timeStepA0Compound}), and we
therefore write $t = \eps t'$, where $t'=O(1)$. These scalings all follow from the observation that
rapid advection will be balanced by diffusion only on appropriately small spatial scales and that the
evolution will then inevitably take place on the short, advective, time scale.

With the above scalings we arrive at the following PDE, dropping primes,
\begin{equation}\label{eq:PDEA}
  A_t= A_{xx} - (1 - \alpha A^2)A_x + \eps \mu A - \sqrt{\eps} A^3, \qquad (x,t) \in (0,\ell) \times (0,\infty).
\end{equation}
This equation is to be solved subject to the Dirichlet boundary conditions
\begin{equation}\label{eq:PDEBCs}
  A(0,t) = \eta, \qquad A(\ell,t) = 0, \qquad t \in [0,\infty),
\end{equation}
starting from a suitably defined initial condition. The linear stability of steady states
$A(x)$ of~\eqref{eq:PDEA} with the boundary conditions \eqref{eq:PDEBCs} is determined
from the eigenvalues $\lambda$ of the linear problem $\mathcal{L}(A) \psi=\lambda \psi$, where
\begin{equation}\label{eq:PDEStability}
\mathcal{L}(A) = \partial^2_x -(1-\alpha A^2)\partial_x + 2 \alpha A \partial_x A + \eps \mu - 3\sqrt{\eps} A^2
\end{equation}
and $\psi(0)=\psi(\ell)=0$. Thus when $\lambda<0$ ($\lambda>0$) the solution $A(x)$ is
stable (unstable) in time. Since translation invariance is absent, there is no zero
eigenvalue. Henceforth we work exclusively with the rescaled model, and therefore
present numerical results for Eqs.~\eqref{eq:PDEA}--\eqref{eq:PDEBCs}, or
equivalently for the corresponding spatial-dynamical system, introduced in
Section~\ref{phaseplane} below.

\section{Bifurcation diagram for the system~\eqref{eq:PDEA}--\eqref{eq:PDEBCs}}\label{phaseplane}

In this section we employ a spatial dynamics formulation of the steady state BVP
and use it to construct a variety of different solutions that together comprise
the bifurcation diagram. The analysis explains not only the behavior shown in
Figure \ref{mubrPDEAlt} but reveals in addition a multitude of other solution types
whose origin is confirmed by numerical solution of the time-dependent BVP~\eqref{eq:PDEA}--\eqref{eq:PDEBCs}.

\subsection{The spatial van der Pol oscillator} \label{sec:VDP}
It is advantageous to interpret the steady state problem associated
with~\eqref{CCmodel} as a second-order van der Pol equation with independent
variable $x$. We refer to the resulting system as a \textit{spatial van der Pol oscillator}.
We set $z_1 = A$ and use the Li\'enard transformation $z_2 = -A^\prime + A - (\alpha/3)A^3$
to obtain the first-order BVP
\begin{align}
\label{vdpAz1}
& z'_1 = - z_2 + z_1 - \frac{\alpha}{3}z_1^3\\
\label{vdpAz2}
& z'_2 = \sqrt{\eps}(\sqrt{\eps}\mu z_1 - z_1^3)\\
\label{eq:ODEBCs}
& z_1(0) = \eta, \quad z_1(\ell) = 0.
\end{align}
We identify steady states of Eqs.~\eqref{eq:PDEA}--\eqref{eq:PDEBCs} with solutions to
the two-point BVP~\eqref{vdpAz1}--\eqref{eq:ODEBCs}
and use its slow-fast structure to interpret the solution branches shown
in Figure~\ref{mubrPDEAlt}. To this end, we temporarily depart from the
corresponding PDE formulation and represent the solutions of the time-independent
BVP \eqref{eq:PDEA}--\eqref{eq:PDEBCs} in the spatial phase plane. For
this purpose we use the term \textit{spatial stability} or simply
\textit{stability} to classify the asymptotic behavior of stationary solutions
to~\eqref{vdpAz1}--\eqref{vdpAz2}. The \textit{PDE stability} of the corresponding
steady states is determined by solving the eigenvalue
problem~\eqref{eq:PDEStability}, as described in Section~\ref{scaling}.

\begin{figure}
\centering
\includegraphics[width=\textwidth]{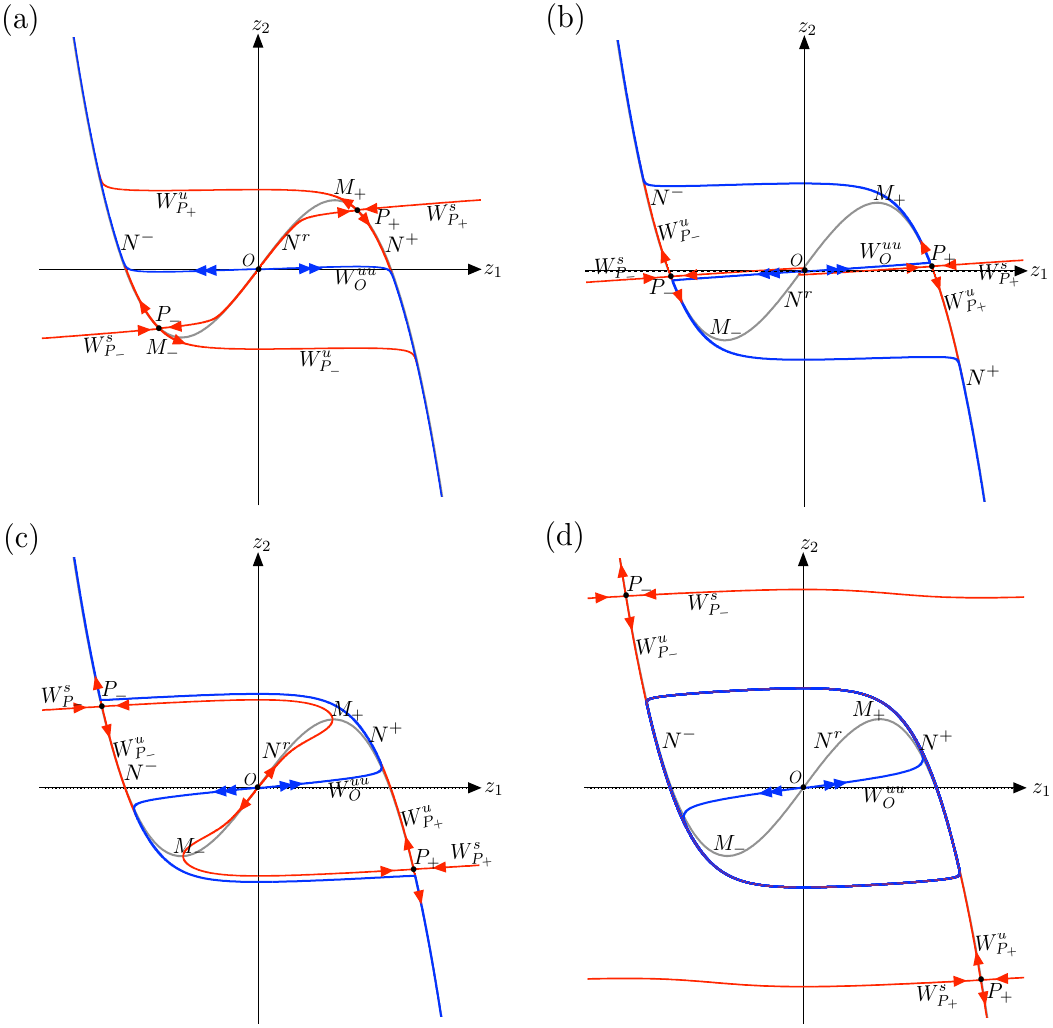}
\caption{Phase planes for Eqs.~\eqref{vdpAz1}--\eqref{vdpAz2} when $\eps=1/144$,
$\alpha=5$, and (a) $\mu=4$, (b) $\mu=7$, (c) $\mu=10$, (d) $\mu=13$.
The phase planes are representative of spatial dynamics in sectors 1--3: (a) sector 1 ($0<\mu<\mu_*$), (b) in sector 2 ($\mu_*<\mu<\mu_{c+}$), close to the transition between 1 and 2, (c)
in sector 2 ($\mu_*<\mu<\mu_{c+}$), close to the transition between 2 and 3, (d) in sector 3 ($\mu>\mu_{c+}$).
For description and definition of $\mu_{c+}$ see Figure \ref{mubrdetail} and
Appendix~\ref{app:pplane}.}
\label{pplane}
\end{figure}
Figure~\ref{pplane} shows the phase plane for this system and various values of the
principal bifurcation parameter $\mu$, with $z_1$ plotted horizontally and $z_2$
vertically. Each panel shows the fast nullcline $N:=N^-\cup N^r\cup N^+$, usually
referred to as the {\it critical manifold} of the slow-fast system,
\begin{equation}
z_2 = z_1 - \frac{\alpha}{3}z_1^3,\label{nullcline}
\end{equation}
together with the line $z_1=0$ along which $z'_2=0$. The critical manifold is divided
into three branches, two (outer) attracting branches $N^{\pm}$ and one (middle)
repelling branch $N^r$. There are three fixed points, one at the origin $O:
(z_1,z_2)=(0,0)$ and the others at
$P_{\pm}:(z_1,z_2)=\pm(\eps^{1/4}\sqrt{\mu},\eps^{1/4}\sqrt{\mu}[1-(\alpha/3)\eps^{1/2}\mu])$.
The former is unstable with eigenvalues
$\lambda=(1/2)[1\pm\sqrt{1-4\eps\mu}]\approx 1,\eps\mu$, while the
latter are always saddles. The eigenvalues of $O$ are real if
$\mu<1/(4\eps)$ but complex if $\mu>1/(4\eps)$. The equality
$\mu=1/(4\eps)$ defines the onset of absolute instability, i.e.,
$\mu=\mu_{\mathrm{a}}$. This is because the linearized two-point BVP
with the boundary conditions $A(0)=A(\ell)=0$, or equivalently
\begin{equation}\label{eq-BVP}
z_1(0)=z_1(\ell)=0,
\end{equation}
only has solutions when the spatial eigenvalues are complex but not otherwise. Note
that in the vicinity of this point the slow-fast decomposition necessarily fails.
However, since this point moves off to infinity in the limit of interest, $\eps\to
0$, this issue does not arise when studying solutions for $\mu=O(1/\sqrt{\eps})$ or
less.

Figure~\ref{pplane} shows that the critical manifold $N$ has a pair of turning points
$M_{\pm}:(z_1,z_2)=\pm(1/\sqrt{\alpha},2/3\sqrt{\alpha})$. Owing to symmetry, in the
following we only consider the saddle $P_+$ and its position on $N$ relative
to $M_+$. We find that for $\mu<1/(\alpha\sqrt{\eps})$
the point $P_+$ lies between $O$ and $M_+$ while for
$\mu>1/(\alpha\sqrt{\eps})$ it lies past $M_+$. In the former case
$P_+$ is attracting along $N$ ($M_+$ is repelling along $N$) while in
the latter case it is repelling ($M_+$ is attracting along $N$; see
Figure~\ref{pplane}). Moreover, the middle part $N^r$ of $N$ is
repelling in the $z_1$ direction while that beyond $M_+$, $N^+$, is
attracting. We also observe that $P_+$ lies on the axis $z_2=0$ when
$\mu=(3/\alpha)\eps^{-1/2}$.\footnote{It is an accident of the
  choice of parameters in \cite{Chomaz99} that this is the case
  precisely at $\mu=\mu_{\mathrm{a}}$.} 
These results suffice to construct the flow of the system in the $(z_1,z_2)$ plane. 
Figure~\ref{pplane} shows the resulting phase plane computed for $\eps=1/144$, 
$\alpha=5$ and several different values of the bifurcation parameter $\mu$. Note in 
particular that panel (a) for $\mu=4$ and panel (c) for $\mu=10$ both show that for 
certain values of $\mu$ the position of the saddle
equilibria $P_{\pm}$ is such that their stable manifold connects (in backward 'time')
to the origin following the middle branch $N^r$ of $N$. As we shall see in the rest
of the paper, this allows for solutions of the BVP~\eqref{vdpAz1}--\eqref{eq:ODEBCs}
for $\eta\neq0$ that also follow $N^r$. Such solutions thus contain canard
segments; see Appendix~\ref{app:pplane} for details.

\begin{figure}
\centering
\includegraphics{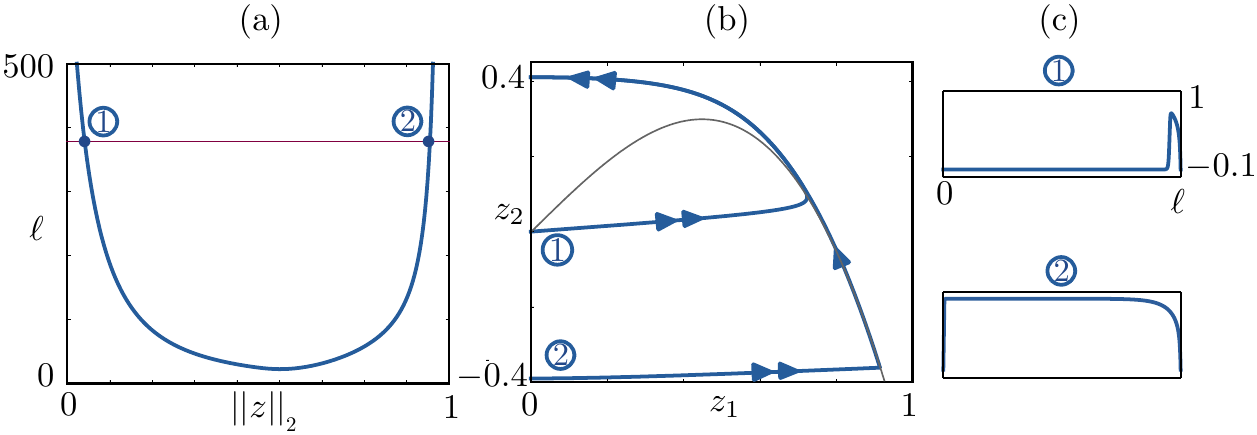}
\caption{(a) Solution branch of Eqs.~\eqref{vdpAz1}--\eqref{eq:ODEBCs} with $\eta=0$,
  obtained by varying
$\ell$ for $\eps = 1/144$, $\alpha = 5$ and $\mu=10$. Two solutions are found for $\ell=377$ 
(straight line), and these are shown as trajectories in the spatial phase plane $(z_1,z_2)$ 
(panel (b)) and as space-dependent stationary states $A(x)$ of the
PDE~\eqref{eq:PDEA}--\eqref{eq:PDEBCs} with $\eta=0$ (panel (c)). 
}
\label{z10twosol}
\end{figure}

We now describe the solution of the BVP (\ref{eq-BVP}), i.e., the
original BVP of \cite{Chomaz97,Chomaz99}, using the
qualitative information contained in Figure~\ref{pplane}.
We assume that $z_1(x)>0$ for $x\in (0,\ell)$. Such solutions exist only
if $z_2(0)<0$. For suppose that $z_2(0)>0$.  Then $z'_1$ at $z_1=0$ is
strictly negative, so that the flow enters immediately into the
negative half plane. Hence solutions of this BVP
can exist only if $z_2(0)<0$. We now argue that such solutions do not
exist if $\mu$ is too small. The critical value of $\mu$ is determined
by the so-called {\it orbit flip} bifurcation at which a branch of
$W^s(P_+)$ coincides with the {\it strong unstable manifold} of $O$,
denoted by $W^{uu}(O)$. As shown by Couairon \& Chomaz \cite{Chomaz97}
this occurs when
$\mu=\mu_{\ast} = 3\alpha^{-1}(U-3\alpha^{-1})\approx 3/(\alpha\sqrt{\eps})$.
For $\mu>\mu_\ast$ the separatrix $W^s(P_+)$ must be below
$W^{uu}(O)$. Hence, continued backwards in `time', it must intersect
the $z_2$ axis at some $z_{2,s}<0$. It follows that for each
$z_2\in (z_{2,s},0)$
there exists $\ell>0$ such that the BVP
\eqref{eq-BVP} has a solution with this choice of $\ell$ and
$z_2(0)=z_2$. Note that $\ell\to\infty$ both as $z_2(0)$ tends to $0$ and
to $z_{2,s}$ (Figure~\ref{z10twosol}). It follows that for $\ell$ large
enough there exist at least two solutions of \eqref{eq-BVP} for the
same $\ell$, one with $z_2$ close to $0$ and one with $z_2$ close to
$z_{2,s}$. The solution with $z_2(0)\approx 0$ spends a long `time'
near $O$ since $O$ is a fixed point, which implies that its $L_2$ norm
is small. The solution starting near $z_{2,s}$ spends a long `time'
near $P_+$ for the same reason and its $L_2$ norm is therefore
large. As $\mu$ approaches $\mu_\ast$ from above these two solutions
approach each other and the solution with the small $L_2$ norm
develops a longer segment near $P_+$, implying rapid growth of the $L_2$
norm. Hence the quasi-vertical segment of the solution branch near
$\mu=\mu_\ast$ (Figure~\ref{mubrPDEAlt}(a)) is due to an orbit flip.
\begin{figure}
\centering
\includegraphics[width=\textwidth]{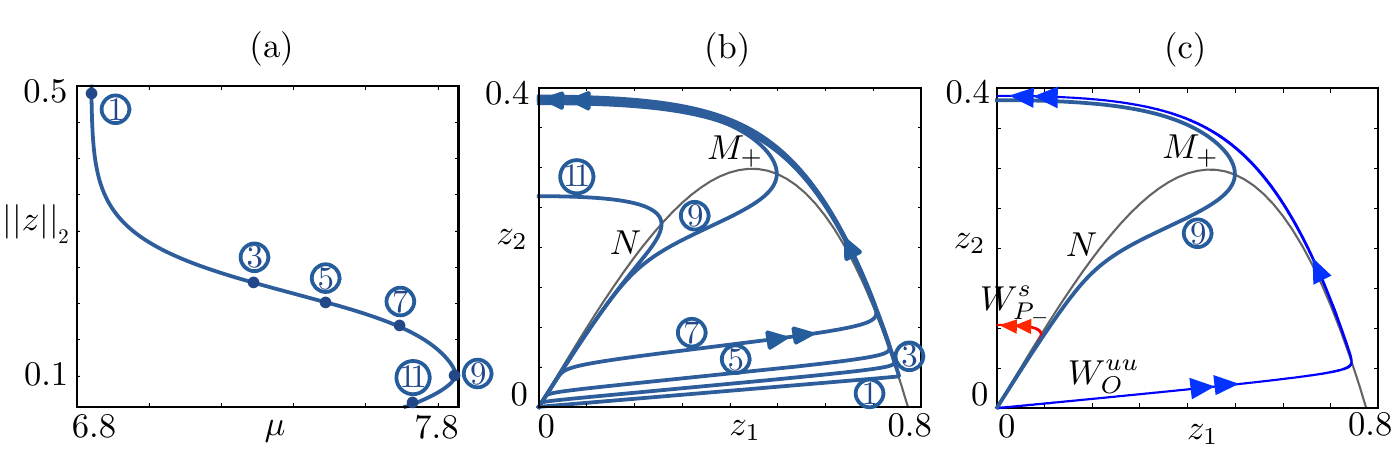}
\caption{(a) Detail of the small amplitude convective instability branch computed
  from~\eqref{vdpAz1}--\eqref{eq:ODEBCs} with $\alpha=5$, $\eps=1/144$ and
$\eta=10^{-10}$. (b) Portion of the phase plane showing the trajectories
corresponding to the points indicated in (a). (c) Trajectory 9 from (a) and (b),
displaying a canard segment, together with some of the invariant manifolds of
equilibria of the system present at this value of $\mu$. The point $P_+$ lies below
the $z_1$ axis. Figure~\ref{mubrdetail}(a) provides are larger view of this case.
\label{canards}}
\end{figure}

For comparison we show in Figure \ref{canards} the corresponding results for
\begin{equation}\label{eq-pnBVP}
z_1(0)= \eta, \qquad z_1(\ell)=0
\end{equation}
with $\eta=10^{-10}$. The figure corresponds to sector 2 (panels (b) and (c) in Figure \ref{pplane})
and reveals the appearance of a new fold in the solution branch (Figure \ref{canards}(a)).
This fold is visible here because of the larger value of $\alpha$ used; such a fold is present
for $\alpha=2$ also but only for even smaller values of $\eta$ than those used in
generating Figure \ref{mubrPDEAlt}. The phase portraits corresponding to the locations 1--11 on
the branch in Figure \ref{canards}(a) are shown in Figure~\ref{canards}(b). Of particular interest
in this paper are trajectories resembling trajectory 9 (Figure~\ref{canards}(c)) which follows for
a 'time' the repelling manifold $N^r$ before reaching $z_1=0$ at a finite value of $z_2$. These
trajectories are associated with the canard phenomenon and are discussed in detail in
Appendix \ref{app:pplane}.


\subsection{Solution profiles}

Figure~\ref{mubrPDEAlt}(b) compares the solution profiles along the
convective and absolute instability branches for $\alpha = 2$, starting with
the profiles 1--3 along the absolute instability branch. These solutions can be
computed for $\eta=0$ since they are insensitive to the precise value of $\eta$. The
location of the solutions on the branch is indicated in
Figure~\ref{mubrPDEAlt}(a). Profile 3 shows a solution that is
strongly compressed against the downstream boundary; this solution
resembles the neutral wall mode predicted by linear stability theory
for the state $A\equiv 0$. Decreasing $\mu$ leads to an increase in
the amplitude of the solution (Figure~\ref{mubrPDEAlt}(a)); at the
same time the solution broadens and starts to fill the domain. Profile
2 corresponds to $\mu=\mu_{\ast}$ and is located in the region of
abrupt amplitude growth associated with the orbit flip. We see that
the solution develops a plateau corresponding to $A=\sqrt{\mu_{\ast}}$
and that the length of this plateau increases as the solution norm
increases. The shape of the front connecting this plateau to $A=0$
remains fixed but the location of the front migrates very rapidly
upstream. This process terminates when the front reaches the upstream
boundary and so becomes pinned at a particular location. Subsequent to
this pinning event the branch turns around to larger values of $\mu$
and the amplitude of the plateau grows in proportion to $\sqrt{\mu}$
(profile 1).

Profiles 4--8 in Figure~\ref{mubrPDEAlt}(b) are instead selected from
the convective instability branches. Profiles 4--6 show the evolution
of the solution with increasing $\mu$ when $\eta=10^{-5}$.
Near $\mu=0$ the solution decreases from $\eta=10^{-5}$ at $x=0$ towards
$A=0$ at $x=\ell$. The slope is almost constant throughout the domain,
indicating a balance between $UA_x$ and $\mu A$ with small adjustment
near the downstream boundary (profile 4). As $\mu$ increases the slope
reverses, with the amplitude growing towards the downstream boundary
(profile 5). As $\mu$ increases further the amplitude near the
downstream boundary saturates and the solution develops a plateau that
evolves in a similar fashion to that on the absolute instability
branch with increasing $\mu$ (profile 6). Profiles 7 and 8 correspond
to branches with yet smaller $\eta$ ($\eta=10^{-9}$ and $10^{-13}$, respectively); 
since decreasing $\eta$ delays the onset of spatial growth the solutions reach
a given norm at larger $\mu$ and hence with a higher amplitude of the plateau.
For yet smaller values of $\eta$ (not shown) one find the behavior shown in
Figure \ref{canards}(a) with an additional fold at $\mu=\mu_{\eta}$.

\subsection{Stability in time}

A standard PDE stability calculation~\cite{Trefethen97} shows that the convective
instability branches are stable in time throughout their range of existence when
the fold at $\mu=\mu_{\eta}$ is absent, and stable in $0<\mu< \mu_{\eta}$ when the
fold $\mu_{\eta}$ is present. In the former case stability is transferred at the
degenerate fold to the upper part of the absolute instability branch; this branch
is also stable. In the latter case there is a hysteretic transition to the upper
absolute instability branch at $\mu=\mu_{\eta}$. Thus a stable steady solution is in fact
present for all $\mu>0$ provided only that $A(0)\ne0$ (Figure \ref{mubrPDEAlt}).

\begin{figure}
\centering
\includegraphics{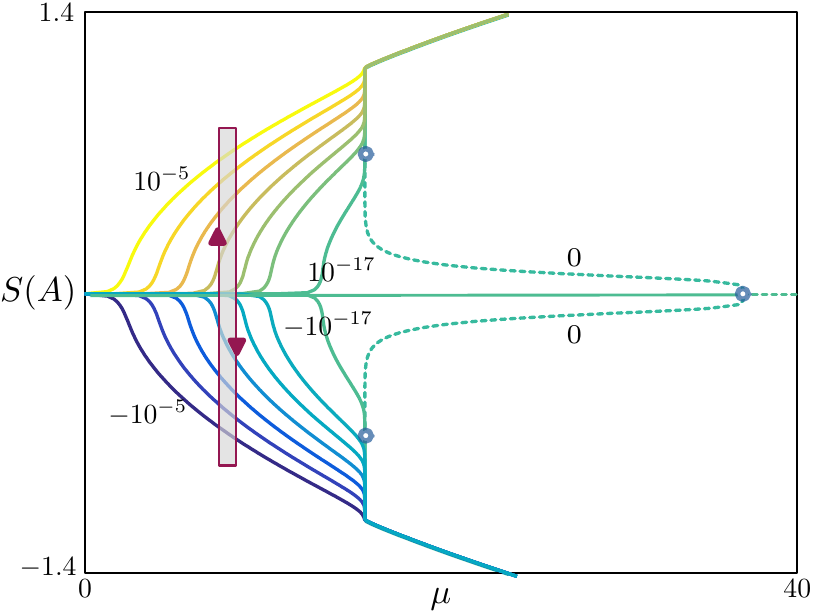}
\caption{Branches of steady states of Eq.~\eqref{eq:PDEA} with $\alpha\!=\!2$ and 
  various inlet boundary conditions $\eta \in [-10^{-5},10^{-5}]$ with
  a superposed cycle (in red) showing changes in the solution
  amplitude when $\eta$ oscillates quasistatically in time with
  $O(10^{-5})$ amplitude and mean $\eta=0$. A concrete example is
  shown in Figure~\ref{fig:A0SinusoidalSweep}.}
\label{fig:PDEBifurcationDiagram_Alpha_2}
\end{figure}

\begin{figure}
\centering
\includegraphics{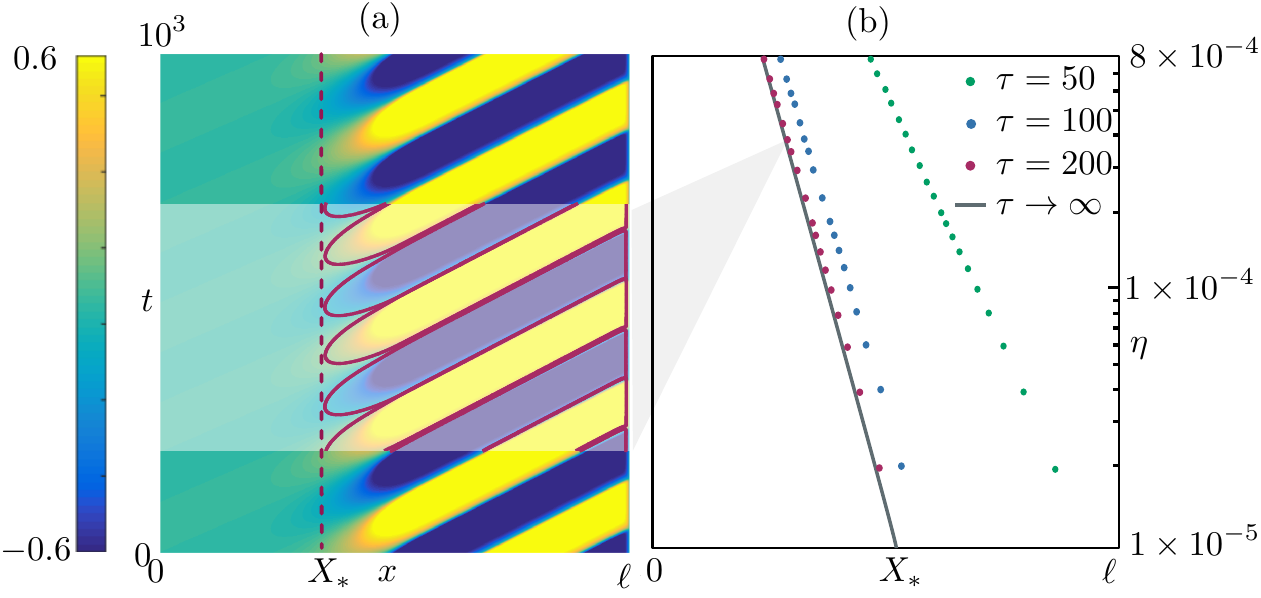}
\caption{(a) Location $X_*$ from Eq.~\eqref{eq:XStar} of time-periodic
fronts when Eq.~\eqref{eq:PDEA} is subjected to a time-periodic inlet boundary
  condition $A(0,t) = \eta \sin(2 \pi t/\tau)$; such fronts are
  sustained entirely by the small amplitude modulation in $A(0,t)$ (in
  the example shown here $\mu=6.1$, $\eta = 4\times 10^{-4}$,
  $\tau = 200$, with the other parameters as in
  Figure~\ref{fig:PDEBifurcationDiagram_Alpha_2}). The red lines in
  the central part are the contour lines $A(x,t) = \pm 0.1$. (b) The
  front location $X_*$ for various values $\eta$ and $\tau$ (dots).
  The steady states of Figure~\ref{fig:PDEBifurcationDiagram_Alpha_2} are continued
  at $\mu=6.07$ in the parameter $\eta$. These states have a well-defined front
  location, so we use the resulting $X_*$ as a solution measure (curve marked $\tau
  \to \infty$). The large $\tau$ time-periodic simulations are well approximated by the
  steady-state continuation.
}
\label{fig:A0SinusoidalSweep}
\end{figure}

\section{PDE simulations}\label{sec:simulations}

Given the sensitivity of the system to the inlet boundary condition
$A(0,t)=\eta$, a question arises as to how solutions behave when this parameter is
time-dependent or randomly distributed.

\subsection{Time-dependent inlet boundary condition}\label{timedeptinlet}

Figure~\ref{fig:PDEBifurcationDiagram_Alpha_2} presents the solution
branches computed from Eq.~\eqref{eq:PDEA} with $\alpha=2$ and different values
of the inlet amplitude $\eta$. The results are similar to those
presented in the previous sections for $\alpha = 5$: as $\eta$ varies
within an $O(10^{-5})$ interval of $\eta=0$ the front location in the
corresponding steady states changes dramatically, as witnessed by the
$O(1)$ variations in the solution measure $S(A)$.

We now fix $\mu$ at a value below $\mu_*$ and prescribe time-periodic inlet boundary
conditions, $A(0,t) = \eta \sin(2 \pi t/\tau)$. As sketched in
Figure~\ref{fig:PDEBifurcationDiagram_Alpha_2}, for fixed $\mu$ and
sufficiently large $\tau$ we expect the front location to oscillate as
the spatio-temporal solution drifts from one branch of stable
stationary states to the next and back again.

In Figure~\ref{fig:A0SinusoidalSweep}(a) we present a numerical experiment for $\mu=6.07$
with a slowly-varying, small-amplitude inlet condition ($\eta = 4\times 10^{-4}$,
$\tau = 200$). As expected, a time-periodic front establishes
towards the center of the domain; we remark that this front is sustained solely
by the small periodic variations in the inlet boundary conditions: if
$\eta = 0$ the only attractive state in this region of parameter space is
the trivial state $A = 0$, as seen in Figure~\ref{fig:PDEBifurcationDiagram_Alpha_2}.

In order to quantify the position of the front, we monitor the level
sets of $A(x,t)$ as $t$ varies. In Figure~\ref{fig:PDEBifurcationDiagram_Alpha_2}
we define this position to be\footnote{Strictly speaking, the minimum is taken only
over the subset of $[0,T]$ for which the threshold $A(x,\cdot)=0.1$ is attained,
for there may be times for which $|A(x,t)| < 0.1$ for all $x \in [0,\ell]$.}
\begin{equation}\label{eq:XStar}
X_* = \min_{t \in [0,T]} \{ x \in [0,\ell] \colon \vert A(x,t) \vert  = 0.1 \}.
\end{equation}
We expect the steady state analysis presented in the previous section to
provide information about the front location, and this is confirmed by
the results in Figure~\ref{fig:A0SinusoidalSweep}. For solutions with a
nonoscillatory profile, such as those considered here, we expect the results to be
robust with respect to changes in the threshold value. The figure shows
the results of steady state continuation in the parameter $\eta$, using
$X_*$ as a solution measure (curve labeled $\tau \to \infty$ in
Figure~\ref{fig:A0SinusoidalSweep}(b)): the solutions on this branch
are steady states, but they have a well-defined $X_*$ which
approximates the location of time-periodic fronts like that in
Figure~\ref{fig:A0SinusoidalSweep}(a). We repeated the direct
numerical simulations for various values of $\eta$ and $\tau$, and
found that the steady-state approximation is excellent for sufficiently slow
oscillations of the inlet boundary conditions (Figure~\ref{fig:A0SinusoidalSweep}(b), $\tau = 200$).
However, for faster oscillations (Figure~\ref{fig:A0SinusoidalSweep}(b), $\tau = 50$,
$100$) the dependence of $X_*$ on $\eta$ remains logarithmic but with a shallower
slope that depends on $\tau$.
\begin{figure}
  \centering
  \includegraphics{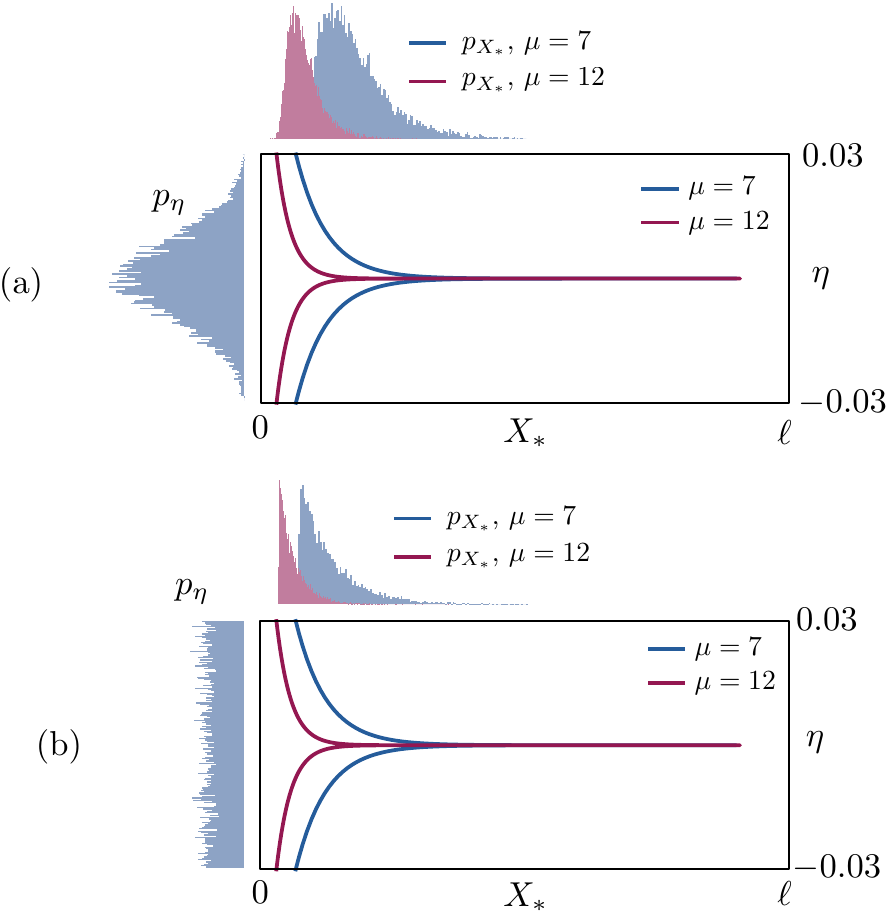}
  \caption{For randomly distributed inlet boundary conditions $\eta$ with density
  $p_{\eta}$, the location $X_*$ of the front in problem~\eqref{eq:UQPDE}
  is distributed with induced density $p_{X_*}$. Since for each sample there is a deterministic
  relation between $\eta$ and $X_*$ (the blue and red branches $X_*(\eta)$), we
  infer the density $p_{X_*}$ directly from $X_*(\eta)$ (upper panels).
  (a) If $p_{\eta}  = \mathcal{N}(0,0.01)$, then $X_*$ is approximately
  Beta-distributed; we observe $\mathbb{E}[X_*] \to 0$, $\var[X_*] \to 0$ as $\mu \to
  \mu_\ast$, i.e., the distribution of front locations moves towards the inlet as
  $\mu$ increases towards $\mu_\ast$. (b) The experiment is repeated, with
  qualitatively similar results, for $p_{\eta} = \mathcal{U}(-0.03,0.03)$. The
  distribution $p_{X_*}$ is now approximately exponential. Parameters are as in
  Figure~\ref{fig:PDEBifurcationDiagram_Alpha_2}.
}
  \label{fig:histogramsPreCritical}
\end{figure}

\begin{figure}
  \centering
  \includegraphics{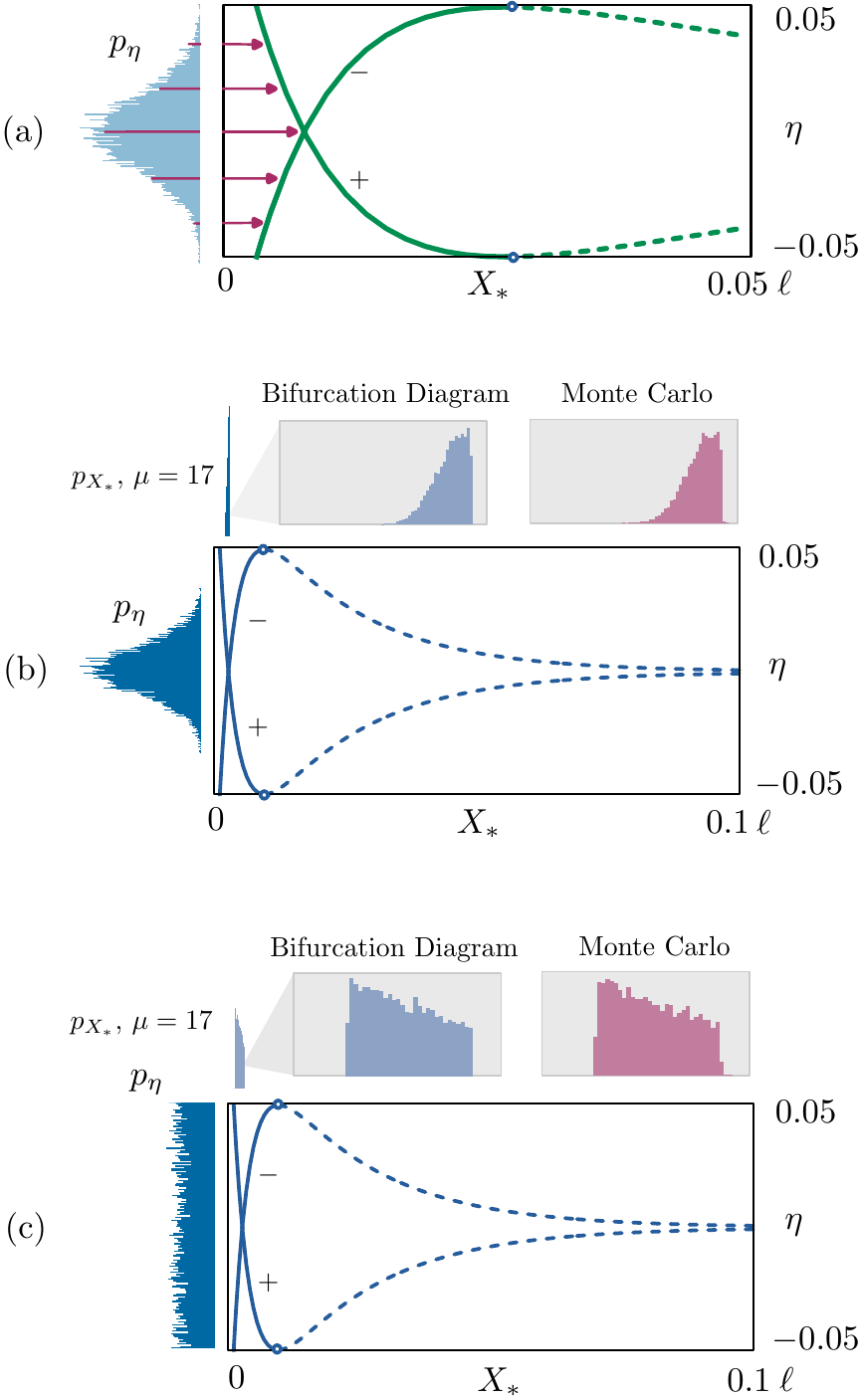}
  \caption{As for Figure~\ref{fig:histogramsPreCritical} but repeated
    for $\mu=17 > \mu_\ast$. (a) At this value of $\mu$ the
    deterministic system possesses two stable solutions that depend on
    the inlet amplitude $\eta$. We label these solutions using the
    symbols $-$ and $+$ according to $A(X_*)>0$ and $A(X_*)<0$,
    respectively.  These solutions remain stable (solid lines) up to a
    maximum value of $\eta$ and are unstable beyond (dashed
    lines). Since $S(A(x,0,\omega)) \approx 0$, we expect that the
    system~\eqref{eq:UQPDE} will evolve towards $X^+_*$ when
    $\eta(\omega) > 0$, and towards $X^-_*$ when $\eta(\omega) < 0$, as
    indicated by the arrows. (b), (c) The hypothesis of panel (a) is
    used to approximate $p_{X_*}$ (upper panels); for $\mu = 17$ the
    front is located very close to the inlet, and the histograms
    compare favorably with Monte Carlo simulations of~\eqref{eq:UQPDE}
    with $T=10^3$ and $10^4$ samples.
  }
  \label{fig:histogramsPostCritical}
\end{figure}

\subsection{Uncertainty quantification of the front location}\label{randominlet}

Motivated by the discussion in the previous sections, we consider steady states of the
problem
\begin{equation}\label{eq:UQPDE}
\begin{aligned}
& A_t= A_{xx} - (1 - \alpha A^2)A_x + \eps \mu A - \sqrt{\eps} A^3, & (x,t) \in (0,\ell) \times (0,T),\\
& A(0,t,\omega) = \eta(\omega), \quad A(\ell,t,\omega) = 0,                           & t \in [0,T], \\
& A(x,0,\omega) = ( 1 - x/\ell)\eta(\omega),                                   & x \in [0,\ell],
\end{aligned}
\end{equation}
where the random inlet parameter
$\eta(\omega) \colon \Omega \to \mathbb{R}$ has an associated density
$p_{\eta}$. In what follows, we will choose $p_{\eta}$ so as to give
samples $| \eta | \ll 1$, that is, we prescribe small random inlet
boundary conditions and study the propagation of uncertainty in the
front location $X_*$ of steady states attained from approximately flat
initial conditions (note that the slope of the initial condition
in~\eqref{eq:UQPDE} is negligible, since $|\eta| \ll 1$ and
$\ell\gg 1$).

Once again, the bifurcation diagrams of Figure~\ref{fig:PDEBifurcationDiagram_Alpha_2}
help us to understand the problem.  We start
by fixing $\mu < \mu_\ast = 15.75$. For a fixed event $\omega \in \Omega$, there is
a single attracting steady state $A(x,\omega)$ with a front at $x=X_*$ that satisfies
$|A(X_*(\omega),\omega)|= 0.1$. Moreover, for fixed $\omega \in \Omega$,
the random variable $X_*$ is a \textit{deterministic} function of $\eta$: the graph of such
a function is approximated by the curve labeled $\tau \to \infty$ in
Figure~\ref{fig:A0SinusoidalSweep}(b). This means that we can infer directly the
density $p_{X_*}$ using $p_{\eta}$ and the bifurcation diagram. In
Figure~\ref{fig:histogramsPreCritical}(a) we use numerical continuation to compute
$X_*(\eta)$ in the interval $\eta \in [-0.03, 0.03]$ for $\mu = 7,12 < \mu_\ast$;
we then prescribe a normally distributed inlet boundary condition
$p_{\eta} = \mathcal{N}(0,0.01)$ and compute the corresponding densities $p_{X_*}$.

The histograms for $p_{X_*}$, obtained using $10^4$ samples, show that
the distribution of the front locations is shifted towards the inlet;
we observe, in addition, that $p_{X_*}$ resembles a Beta distribution
with $\mathbb{E}[X_*] \to 0$, $\var[X_*] \to 0$ as $\mu \to
\mu_\ast$.
Consequently, the uncertainty in the front location decreases
considerably as we transition towards $\mu_\ast$: for $\mu = 7$, some
samples have fronts close to $\ell/2$, as can be seen in the blue
histogram of Figure~\ref{fig:histogramsPreCritical}(a), while the
distribution for $\mu = 12$ peaks much closer to the inlet. In
Figure~\ref{fig:histogramsPreCritical}(b) we repeat the calculations
for uniformly distributed inlet boundary conditions,
$p_{\eta} = \mathcal{U}(-0.03,0.03)$. The results are qualitatively
similar to the normally distributed case, except that $p_{X_*}$ is now
approximated by an exponential distribution.

A complication arises when we consider the case $\mu > \mu_\ast = 15.7$. In
this region of parameter space more than one attracting steady state may be
present for each fixed event $\omega$. We have not shown such branches in
Figure~\ref{fig:PDEBifurcationDiagram_Alpha_2}, but refer the reader to
Section \ref{sec:disc}, Figure~\ref{mubrdetail}(a), for an example.

Figure~\ref{fig:histogramsPostCritical} shows the results for
$\mu=17 > \mu_\ast$, i.e., in a region of parameter space where two
stable solutions with different $X_*$ coexist. Both branches are
initially stable (solid lines) and destabilize at saddle-node
bifurcations; importantly, there are two stable attracting steady
states for each $\eta \in [-0.05,0.05]$, selected by the initial
conditions for the problem~\eqref{eq:UQPDE}. Since
$S(A(x,0,\omega)) \approx 0$, it is reasonable to expect that the
system will be attracted to a solution on the branch with $A(X_*)>0$
when $\eta(\omega) <0$ (labeled $-$), and to a solution on the branch
with $A(X_*)<0$ when $\eta(\omega) >0$ (labeled $+$), as sketched in
Figure~\ref{fig:histogramsPostCritical}(a). In passing, we note that
the $X_*$ scale in Figure~\ref{fig:histogramsPostCritical}(a)
indicates that the front location is much closer to the inlet,
$X_* \in [0,0.05\ell]$, for $\mu > \mu_\ast$ than when $\mu < \mu_\ast$
(cf. Figure~\ref{fig:histogramsPreCritical}).

The above conjecture is supported by the numerical experiments reported in
Figures~\ref{fig:histogramsPostCritical}(b,c). We repeated the procedure of
Figure~\ref{fig:histogramsPreCritical} for $\mu = 17$ assuming that
\[
X_*(\eta) =
\begin{cases}
  X^+_*(\eta) & \text{if $\eta < 0$,} \\
  X^-_*(\eta) & \text{if $\eta \geq 0$}
\end{cases}
\]
and found that the resulting densities $p_{X_*}$ are sharply peaked
around the inlet (upper panels in (b,c)). The insets show an excellent
agreement with the histograms computed via Monte Carlo simulations of
the system~\eqref{eq:UQPDE} using $T=10^3$ and $10^4$ samples.

\subsection{Stochastic simulations}\label{stochasticinlet}
\begin{figure}
  \centering
  \includegraphics[width=\textwidth]{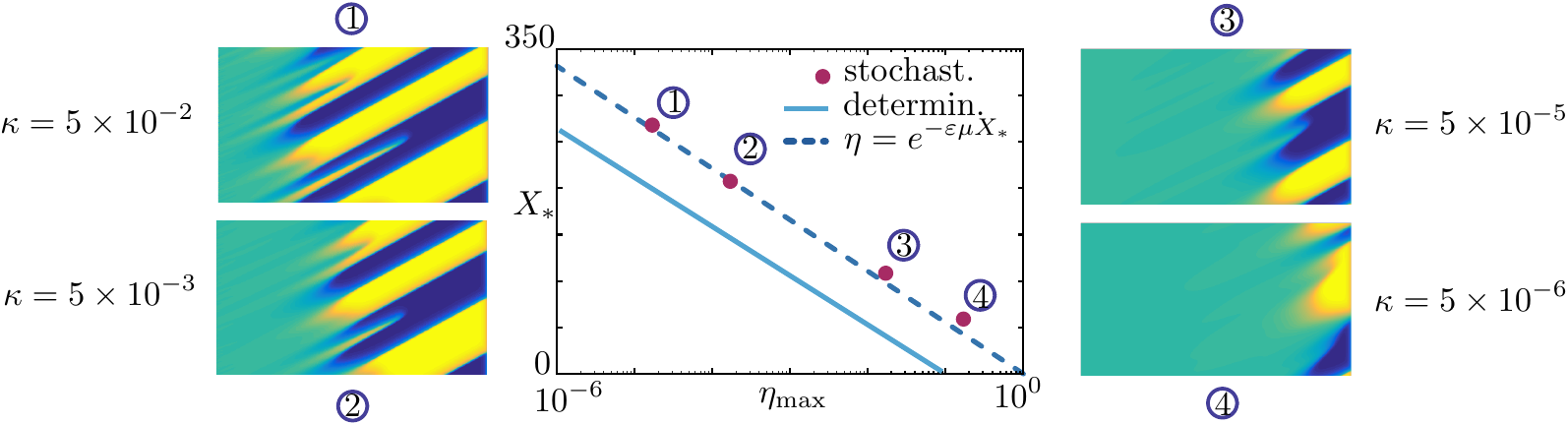}
  \caption{Direct numerical simulations of Eq.~\eqref{eq:PDEA} with a
    stochastic inlet boundary condition $A(0,t) = \kappa \tilde
    \eta(t)$, where $\kappa > 0$ and $\tilde \eta(t)$ is a Wiener process (see
    Eq.~\eqref{eq:Wiener}). We show contour plots of $A(x,t)$ for
    $(x,t)\in[0,377]\times[0,500]$ for $\mu=6$, $\eps = 1/144$ and
    various values of $\kappa$ (panels 1--4). 
    When $\kappa = 0$ the problem is deterministic and the
    only attracting solution is the trivial steady state $A\equiv 0$
    but, as the noise is turned on ($\kappa >0$), noise-sustained structures
    appear. As $\kappa$ increases 
    so does the maximum inlet amplitude $\eta_\mathrm{max} = \max_t | \eta(t) |$,
    resulting in decreased distance $X_*$ to the noise-sustained
    front.  The deterministic and stochastic setting have a common
    scaling $\eta = e^{-\eps\mu X_*}$ (see text).}
  \label{noise}
\end{figure}

We report, finally, on the results of stochastic simulations in which
the inlet value is a continuous stochastic process. Specifically, we set
$A(0,t) = \eta(t) = \kappa \tilde \eta(t)$, where $\kappa > 0$ and $\tilde
\eta(t)$ is a Wiener process, i.e.,
\begin{equation}
\tilde \eta(0) = 0 \text{ a.s, } \qquad \tilde \eta(t+s) - \tilde \eta(t) \stackrel{\text{i.i.d.}}{\sim} \sqrt{s} \,\mathcal{N}_{0,1}, \quad \text{for all $t,s > 0$}.
\label{eq:Wiener}
\end{equation}
In the following we use the standard Euler-Maruyama scheme~\cite{Lord14} to
integrate the resulting stochastic PDE. The results in
Figure~\ref{noise} demonstrate the presence of noise-sustained
structures in the regime $0<\mu<\mu_{\ast}$, thereby generalizing the
results for fixed but randomly selected values of $\eta$ in Section
  \ref{sec:simulations}\ref{randominlet}
  and those for periodically oscillating inlet boundary conditions in
  Section \ref{sec:simulations}\ref{timedeptinlet}. Since the only source of
  randomness is in the boundary condition $A(0,t) = \kappa \eta(t)$, by setting
  $\kappa=0$ we recover the deterministic system with $0 < \mu < \mu_*$ and $A(0,t) =
  0$ for all $t$. The only attracting solution of this system is the trivial
steady state $A = 0$.
As the noise is turned on, noise-sustained structures appear, echoing
the behavior found for random but time-independent inlet conditions
(Figure \ref{fig:A0SinusoidalSweep}(b)). In order to make a
quantitative comparison with the deterministic case, we compute $X_*$
using the definition~\eqref{eq:XStar} and plot this value against
$\eta_\textrm{max} = \max_t |\eta(t)|$ for each of the realizations reported in
Figure~\ref{noise}. Here $\eta_\textrm{max}$ is used as a proxy for the noise strength.

The $X_*(\eta_\textrm{max})$ scaling
observed in Figure~\ref{noise}
can be understood completely in terms of the deterministic case 
(solid blue line in Figure~\ref{noise}): this curve, obtained by numerical
continuation, is the semi-logarithmic plot of the $X_*(\eta)$ curve in
Figure~\ref{fig:A0SinusoidalSweep}(b) for $\mu=6$. To each
deterministic stationary profile $A(x)$ there corresponds an orbit of the
spatial dynamical system~\eqref{vdpAz1}--\eqref{eq:ODEBCs} with initial condition
close to the origin $(0,0)$.  As noted in Section~\ref{sec:simulations}(a), the origin is
unstable with eigenvalues
$\lambda_{1,2} \approx \eps \mu, 1$. The profiles $A(x)$ of interest (see, for
instance, panels $6$, $7$, $8$ of Figure~\ref{mubrPDEAlt}) correspond to orbits
that spend a long ``time'' close to the origin, following the repelling slow
manifold: these orbits drift along the weak eigendirection of the origin, before being
ejected and crossing the threshold $|A(x)| = 0.1$. Since
\[
A(x) \approx \eta e^{\eps\mu x}, \qquad x \in [0,0.1],
\]
it follows that $X_*$ is determined by the condition $0.1\approx \eta e^{\eps\mu X_*}$
which, as seen in Figure~\ref{noise}, represents the common scaling between the stochastic
and deterministic cases to good accuracy.

\section{Discussion} \label{sec:disc}

In this paper we have examined the properties of a prototypical system
describing a subcritical pattern-forming instability in the presence
of strong advection.  Problems of this type are usually formulated
with Dirichlet boundary conditions at both the inlet and outlet
locations. In the case $\eta=0$ (zero inlet boundary condition)
the base flow loses stability at the threshold
for the first appearance of a global mode, $\mu=\mu_\mathrm{g}$, a
parameter value close to that for the onset of absolute instability,
$\mu=\mu_{\mathrm{a}}$, when this system is formulated on the real line
instead. We have described the consequences of
nonlinear absolute instability that is responsible for the presence of
a nonlinear global mode for values of the parameter $\mu$ in the range
$\mu_{\ast}<\mu<\mu_\mathrm{g}$,
where $\mu_\mathrm{g}$ corresponds to the subcritical
bifurcation of the trivial state, and $\mu_*$ to the fold bifurcation where the
subcritical branch restabilizes (see Figure~\ref{fig:chomazResults}). Thus no
persistent states are
present for $\mu<\mu_{\ast}$, and our analysis showed that the fold at
$\mu=\mu_*$ is due to the presence of an orbit flip in the
spatial dynamics description of the system and hence is highly
degenerate. The spatial dynamics formulation also led us to the
identification of a new set of nontrivial states present in the
range $0<\mu<\mu_\mathrm{g}$ (i.e., in the convectively unstable
regime) whenever $\eta\ne0$ (i.e., for nonzero inlet boundary conditions).
These new states are a consequence of canard segments identified
in the phase plane analysis and these in turn are responsible
for the extreme sensitivity of the resulting steady states
to the details of the inlet boundary condition. Direct numerical
simulation of the PDE model allowed us to associate these states
with the noise-sustained structures familiar from numerical studies
of systems of this type in the convectively unstable regime
\cite{Deissler87}. This sensitivity is in stark contrast with the role
of the outlet boundary condition which has essentially no influence on
the behavior described here~\cite{Chomaz99,Tobias98}. This is because all
infinitesimal, spatially localized, perturbations with upstream phase
velocity are necessarily evanescent (in space). Thus the influence of
the downstream boundary conditions extends but a short distance upstream
of the outlet boundary condition, and the behavior in the bulk of the domain
remains unchanged.
\begin{figure}
\centering
\includegraphics[width=\textwidth]{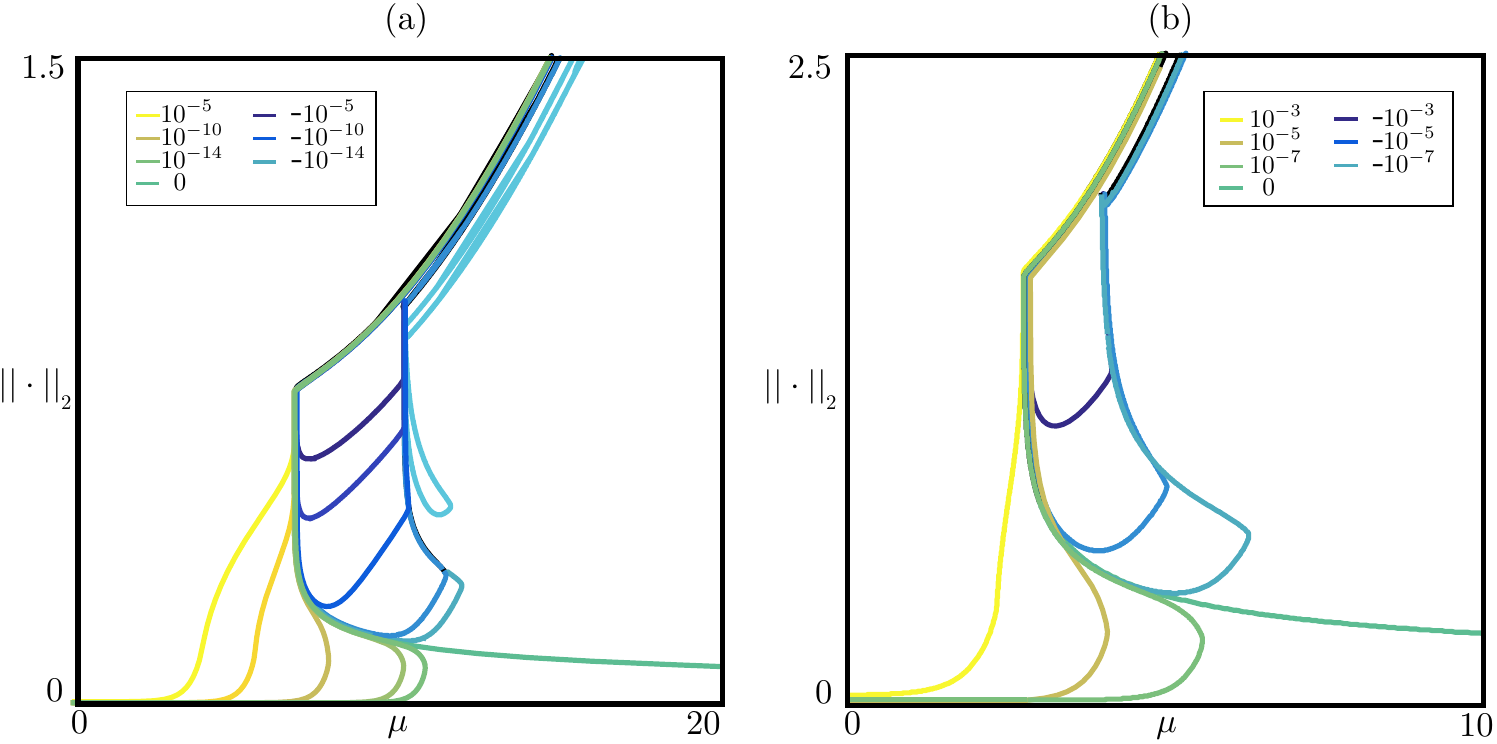}
\caption{Solution branches for different values of $z_1(0)\equiv\eta$ when
$\alpha=5$, $\eps=1/144$ in system~\eqref{vdpAz1}--\eqref{vdpAz2} (panel (a)), and
when $\alpha=1$, $\eps=1/144$ in system~\eqref{vdpBz1}--\eqref{vdpBz2} (panel (b)).
The panels show branches for $z_1(0)=\eta\ne0$; these differ in the sign of $\eta$
(see legend) and are organized by an orbit-flip bifurcation at $\mu=\mu_{\ast}$ (see
text). The origin of the additional branches revealed in this figure, including the
second vertical segment at $\mu=\mu_{c+}$, is discussed in Appendix
\ref{app:pplane}.
}
\label{mubrdetail}
\end{figure}

It is important to mention that other scalings of our model system \eqref{CCmodel} are
possible. Of these, the scaling
\[
0 < \eps = U^{-2} \ll 1, \quad \alpha \to \eps^{-1/2} \alpha, \quad x \to \eps^{1/2}
x, \quad t \to \eps t,
\]
leads to the spatial dynamical system
\begin{align}
\label{vdpBz1}
z'_1 &= -z_2 + z_1 - \frac{\alpha}{3} z_1^3\\
\label{vdpBz2}
z'_2 &= \eps(\mu z_1 - z_1^3).
\end{align}
In this scaling it is easier to discuss canards (see Appendix~\ref{app:pplane}), but
the bifurcation structure of systems
\eqref{vdpAz1}--\eqref{vdpAz2} and \eqref{vdpBz1}--\eqref{vdpBz2} is
very similar, as confirmed by the bifurcation
diagrams in Figure \ref{mubrdetail}.  Indeed, the rescaling leading to
Eqs.~\eqref{vdpBz1}--\eqref{vdpBz2} demonstrates the robustness of the
behavior we have described here with respect to changes in the
physical parameters of the system.

The problem as formulated here contains two independent parameters,
the forcing parameter $\mu$ responsible for the primary instability
and the advection speed $U$. Similar situations arise in
Rayleigh-B\'enard convection with throughflow. This problem has been
studied by a number of authors, both for transverse rolls
\cite{muller92} and longitudinal rolls \cite{nicolas12}.  However, in
both these cases the pattern-forming instability is supercritical.
This is also the case for Taylor vortices in a Taylor-Couette system
with axial flow \cite{Buchel96,recktenwald93}. Related work on binary fluid
convection \cite{buchel00} studies a subcritical system but with
periodic boundary conditions, thereby eliminating the possible
influence of inlet perturbations.  The effect of such perturbations is
usually studied within a linearized theory \cite{Deissler87} or via
numerical simulations, either of model equations
\cite{Deissler87,Deissler87b,Colet99} or the full partial differential
equations \cite{jung01,nicolas12}. However, in no case has the full
bifurcation diagram been computed for nonzero inlet boundary conditions
as done here.

In contrast, in shear flow instability, including plane Poiseuille
(channel) flow and circular Poiseuille (pipe) flow, the shear flow is
simultaneously responsible for the presence of finite amplitude
instability of the laminar state and the advection of the resulting
turbulent structures downstream \cite{Deissler87c,Eckhardt07}. Such
flows are therefore fundamentally different from the problem studied
here, and indeed in carefully controlled pipe flow experiments the
laminar state persists to large values of the Reynolds number
($Re\sim 60\,000$ \cite{Eckhardt07}), the parameter that
simultaneously quantifies both the shear rate and the advection
velocity, despite the inevitable presence of inlet fluctuations. For
comparison, finite amplitude perturbations trigger pipe flow
turbulence at $Re\sim 2\,000$.

We mention that Eq.~(\ref{CCmodel}), in the form
$A_t=A_{xx}-(U-\alpha |A|^2)A_x+\mu A-|A|^2A$, may also be considered
to be an amplitude equation for the complex amplitude $A(x,t)$ of a
wavetrain of the form $A(x,t)\exp ikx$. The real subspace of this
equation is invariant and identical to Eq.~(\ref{CCmodel}). The
stationary solutions to this BVP constructed here therefore remain
solutions of this higher-dimensional system; the zeros of $A(x)$ now
correspond to phase defects -- locations where the phase of the
solution jumps by $\pi$, and it is possible that the breakup of a
wavetrain into domains with different phase speeds observed in
simulations of the complex Ginzburg-Landau equation with drift
\cite{Gonpe13,Tobias98} can be understood along these lines.

In summary, we expect that the ideas presented in this work may find application in
different areas of fluid mechanics where a strongly subcritical instability interacts
with an imposed flow.

\appendix
\section{Phase plane analysis} \label{app:pplane}

\begin{figure}
\centering
\includegraphics{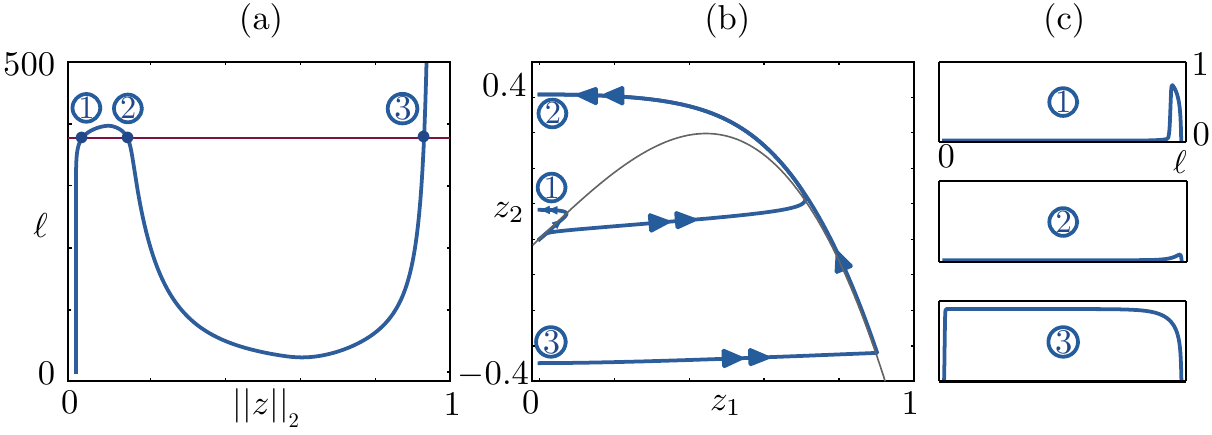}
\caption{(a) Solution branch of Eqs.~\eqref{vdpAz1}--\eqref{vdpAz2} obtained by
  varying $\ell$ for fixed $\mu=10$, $\alpha = 5$ and $\eta=10^{-13}$ for comparison
with Figure \ref{z10twosol}.
  Three solutions are found for $\ell=377$, shown in panels (b) and (c).\label{threesol}
}
\end{figure}
In this section, we refer to convective (absolute) instability states as type I (type
II) solutions. As explained in Section~\ref{sec:VDP}, BVP
\eqref{vdpAz1}--\eqref{eq:ODEBCs} with $\eta = 0$ has a pair
of solutions for $\mu>\mu_\ast$, one of which disappears when its
$L_2$ norm vanishes. This occurs, by definition, at
$\mu=\mu_\mathrm{g}$. Since $\mu_\mathrm{g}\approx \mu_{\mathrm{a}}$
for large $\ell$ we conclude that these solutions are fundamentally a
consequence of absolute instability of the trivial state $A = 0$ that
takes place at $\mu=\mu_{\mathrm{a}}$ and hence of type II. Both solutions start at
$z_2(0)<0$ and follow the fast dynamics to the vicinity of the
critical manifold $N$ above $P_+$, then drift along $N$ past $M_+$,
and finally return to the $z_2$ axis with $z_2(\ell)>0$ (see Figure
\ref{z10twosol}(b)). In the following we refer to solutions for which
$z_2(0)\approx 0$ as type IIa, and to solutions for which the point
$(0, z_2(0))$ is close to $W^s(P_+)$ as type IIb (we assume $\ell$ is
large). Note that no type II solutions are present for $\mu<\mu_\ast$.
The above statements can be made rigorous in the setting of system
~\eqref{vdpBz1}--\eqref{vdpBz2} and explain the transition between
sectors 1 and 2 in Figure~\ref{pplane} and the role of the parameter value $\mu=\mu_*$.

We now consider BVP \eqref{vdpAz1}--\eqref{eq:ODEBCs} with $\eta \neq 0$,
which introduces a small but generic perturbation of the $\eta = 0$ case
that results in a new class of solutions (Figures \ref{mubrPDEAlt} and \ref{canards}).
We think of the boundary
condition $0< |\eta| \ll 1$ as reflecting the presence of small amplitude
imperfections or ``noise'' at the inlet. This perturbation has a number of
consequences. It results in the presence of a nontrivial type I state, and
introduces a reconnection between this state and the type IIa branch very
close to $\mu=\mu_\mathrm{g}$ via a fold bifurcation at $\mu=\mu_{\eta}$
(Figure \ref{canards}(a)); the location
of this fold is extremely sensitive to the value of $\eta$. Branch IIa then continues to
larger amplitude and smaller $\mu$ towards the limit point bifurcation
at $\mu_{\ast}$ (we ignore here the very weak dependence of
$\mu_{\ast}$ on $\eta$) where it connects to the upper IIb branch. It
follows that for the perturbed BVP, with $|\eta|$ very
small, there is an interval $[\mu_{\ast}, \mu_{\eta}]$ of $\mu$
values, with $\mu_{\eta}<\mu_\mathrm{g}$ ($\mu_{\eta}\to\mu_g$ as $\eta\to0$), 
for which both solutions of type II remain (see Figures \ref{z10twosol} and
\ref{mubrdetail}(a)). Since type I solutions are also present, it follows
that for any fixed $\mu\in[\mu_{\ast},\mu_{\eta}]$ there are three
different solutions to the perturbed BVP as illustrated in
Figure~\ref{threesol} (as before, we take $\ell$ large). 
Moreover, for each $\eta>0$, no matter how small, there exists a solution of the perturbed
BVP with $z_2(0)>0$ defined on the interval $\mu\in (0, \mu_{\eta})$.
These type I solutions (Figure \ref{canards}(c)) contain segments that follow $N^r$
but ultimately connect to an attracting fast fiber and so resemble the \textit{faux canards} familiar from the ODE setting, a term used to refer to solutions that start on a repelling slow manifold and
  connect to an attracting slow manifold~\cite{Mitry17}. Other
  notable solutions in slow-fast ODEs are \textit{jump-on canards},
  which start along an attracting fast fiber and connect to a repelling
  slow manifold~\cite{deM2011}. Type I solutions feature this behavior in reverse 'time'.

\begin{figure}
\centering
\includegraphics[width=\textwidth]{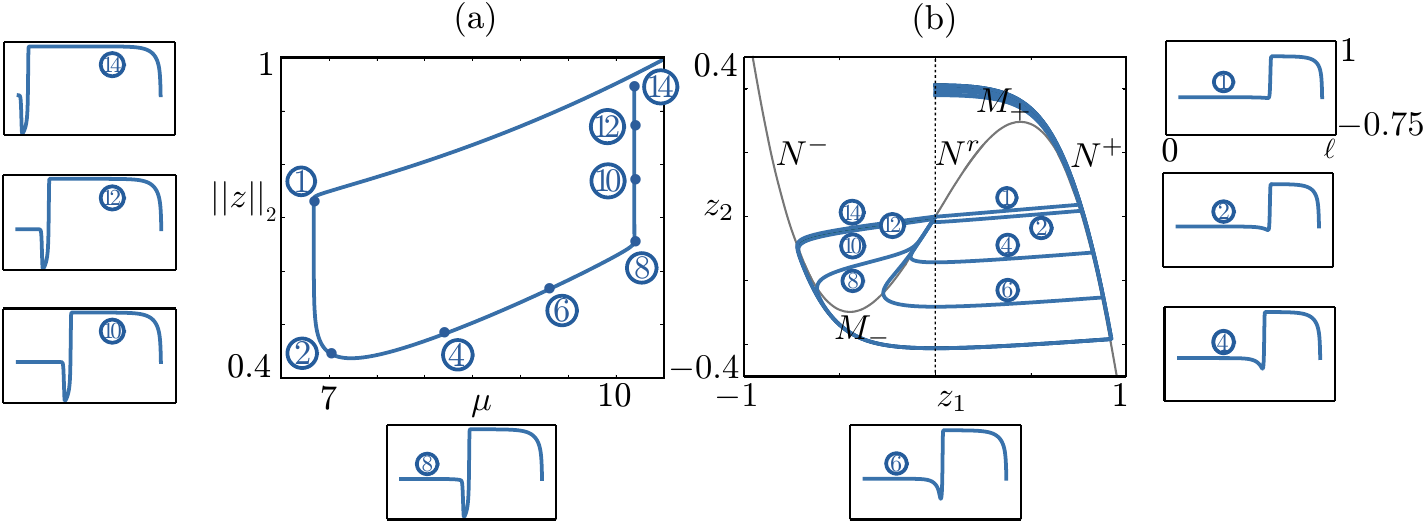}
\caption{(a) Detail of the type I solution branch with $\eta=-10^{-7}$ from
Figure~\ref{mubrdetail}(a). (b) Portion of the phase plane showing the trajectories
corresponding to the locations indicated in (a). Spatial profiles of the solutions of
BVP~\eqref{vdpAz1}--\eqref{eq:ODEBCs} with $\eta < 0$ corresponding to the
trajectories shown in (b) are shown alongside. For the parameter values of
Figure~\ref{mubrdetail}(a)
$\mu_{\ast}\approx 6.84$ and $\mu_{c+}\approx 10.19$.
\label{morecanards}}
\end{figure}

The phase plane plots show that, initially, when $\mu$ is small,
these solutions return very quickly to the $z_1$ axis and their canard
segments are very short. However, as $\mu$ increases so does the length
of the canard segment. For $\mu$ close to $\mu_{\eta}$
the solution of type IIa also develops a canard segment and finally at
$\mu=\mu_{\eta}$ the solutions of type I and type IIa merge in a
limit point bifurcation at $\mu_{\eta}$. For $\mu>\mu_{\eta}$
there exists only one solution, corresponding to type IIb. Note that
for $\mu\in (\mu_\ast,\mu_{\eta})$ there exist three distinct
solutions (see Figure~\ref{mubrdetail}(a) for the smallest values of
$\eta$). The presence of the limit point $\mu_{\eta}$ is thus
related to the transition from solutions analogous to canard cycles 
with no head to ones analogous to canard cycles with a head
(see Figure~\ref{morecanards} and \cite{deMDer}). For larger $\eta$ the
limit point $\mu_{\eta}$ disappears and solutions of type I change
by homotopy to solutions of type II.
 %
%

{\bf Case $\eta>0$}: Figure \ref{canards}(a) shows a detail of the branch corresponding to
$\eta=10^{-10}$ in Figure~\ref{mubrdetail}(a). The numbers 1--11 refer
to points along this branch and the corresponding phase portraits are
shown in Figure~\ref{canards}(b). Note that in contrast to the canard
explosions that occur in exponentially small intervals in
two-dimensional systems \cite{Benoitetal81}, in the present case no
canard explosion takes place and canard behavior occurs over an
$O(1)$ interval of values of the parameter $\mu$.
The absence of an explosive
passage through the canard family is due to the nature of the BVP we consider: we
fix $\ell$, $z_1(0)$ and $z_1(\ell)$. Solutions to this BVP compensate changes in $\mu$ by
adjustments of $z_2(0)$, which controls the distance between $(z_1(0), z_2(0))$ and
the repelling slow manifold. A nearby solution containing a canard segment adjusts
the distance  between $(z_1(0), z_2(0))$ and the repelling manifold, so that the
passage time for the adjusted $\mu$ value is $\ell$: since the passage time along this
manifold does not vary much with $\mu$, the updated solution has a canard segment as well.
Usually canards are computed as solutions to a different BVP, where $z_2(0)$
is fixed and $\ell$ is a free parameter. In this case solutions have canard segments
only for a small range of $\mu$, near the value defined by $(z_1(0), z_2(0))$ on a
repelling slow manifold, and hence explosive behavior occurs.

\begin{figure}
\centering
\includegraphics{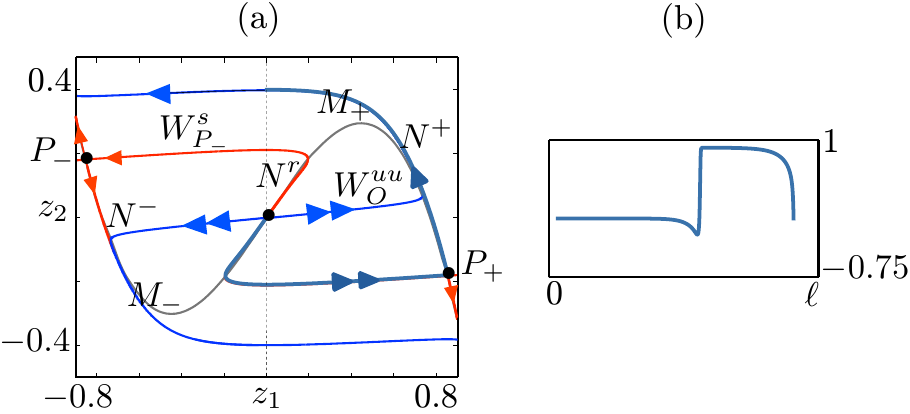}
\caption{(a) Trajectory 5 (thick gray-blue line), containing a canard segment, from
Figure~\ref{morecanards}; also shown is a portion of the phase plane showing the
repelling and attracting nullclines as well as stable and unstable manifolds of the
three equilibria of the system present at this value of $\mu$. (b) Spatial profile of
the corresponding solution to the perturbed BVP~\eqref{vdpAz1}--\eqref{eq:ODEBCs}
with $\eta<0$.
}
\label{canard}
\end{figure}

{\bf Case $\eta<0$}: We next consider the perturbed BVP with $\eta<0$.  Associated solution
branches are shown in Figure~\ref{mubrdetail}(a) and in close-up in
Figure~\ref{morecanards}, which shows the corresponding solution
profiles. Below we describe the solutions labeled $0,1,\ldots$ in
Figure~\ref{morecanards}. Solution labeled 0 is very similar to a type
IIb solution of the unperturbed BVP: it starts close to an unstable fiber
of $N$, jumps to $N^+$, follows it to $M_+$ and
returns to the $z_2$ axis along a fast fiber. As $\mu$ decreases
towards $\mu_\ast$ and $P_+$ moves up along $N$, all solutions are of
this type. This is very similar to the situation for $\eta >0$, since segments of
these branches (for
$\mu>\mu_{\ast}$) remain close to the corresponding solution branches
IIa,b in the unperturbed BVP. The vertical segment between labels 1 and
2 corresponds to a passage near the orbit flip transition and along a
segment of the branch of solutions of type IIb. The branch is similar
to the branches of solutions for $\eta >0$ as it follows the
unperturbed solution branch corresponding to solutions of type IIb for
$\mu$ up to a certain value that increases as $\eta$
decreases. However, the relevant solutions with
$\eta<0$ differ from the corresponding solutions with $\eta>0$ in that
they leave the vicinity of the unperturbed branch ($z_1(0)=0$) in the
opposite direction (see Figure~\ref{mubrdetail}(a)). After that, the
solutions start developing a canard segment and the branch has a fold
that corresponds to the maximal canard segment. The main
difference from the branches with $\eta > 0$ is that
canard segments with $z_1(0)=-\eta$ follow $N^r$
to the left of the vertical axis; compare solutions 2 to 8 in
Figure~\ref{morecanards} with
solutions 9 to 11 in Figure~\ref{canards}. Past the fold point, solutions with
$\eta<0$ containing a canard segment make an excursion towards
$N^-$ before jumping at $M_-$ and following
$N^+$ towards $M_+$; see solutions 8 to 14 in
Figure~\ref{morecanards}.
For the range of $\mu$ values where this
occurs, the saddle point $P_+$ lies below $M_-$. As $\mu$ approaches
the value of $\mu_{c+}$, which corresponds to a connection from the
maximal canard to $P_+$ (see Figures \ref{morecanards}(b) and
\ref{canard}) the corresponding solutions, after jumping near $M_-$,
come closer and closer to $P_+$, which incurs a large increase of the
$L_2$ norm for a very small change in $\mu$ (of the order of $10^{-3}$). This argument explains
the presence of the second vertical part of the branch seen in
Figure~\ref{morecanards}(a) at $\mu=\mu_{c+}$. We mention that
$\mu_{c+}$ is extremely close to the value $\mu=\mu_h$ at which there
is, in the limit $\eps\to0$, a heteroclinic orbit connecting $P_+$ to
$P_-$ via the turning points $M_{\pm}$.  However, the latter solutions
are of no relevance to any of the BVP considered in the present work.
%

We may think of the type I solutions as small amplitude solutions and
type II solutions as large amplitude solutions. The former originate
in convective instability that sets in at $\mu=0$ in the absence of
boundaries while the latter arises from absolute instability. Both
solutions exist when $P_+$ lies below the $z_1$ axis but as $\mu$
decreases and $P_+$ moves above the axis type II solutions are
destroyed via a special solution that follows the strong unstable
manifold of the origin to $P_+$ \cite{Chomaz97}. Solutions leaving $O$
along this manifold are characterized by very rapid increase in
amplitude, followed by a lengthy interval at amplitude
$z_1\approx\sqrt{\mu}$ before decaying back to $z_1=0$ at $x=\ell$. This
condition determines the condition for an orbit flip; as already
mentioned, in the limit $\eps\rightarrow0$, this condition is
equivalent to the passage of $P_+$ through the $z_1$ axis, i.e., to
$\mu=(3/\alpha)\eps^{-1/2}$. When $\alpha=O(1)$ the location of this
point moves to large values of $\mu$; however, in system~\eqref{vdpBz1}--\eqref{vdpBz2}, this
transition occurs at a finite value of $\mu$, $\mu=3/\beta$. In fact,
in the definition of type I/type II solutions we can replace the $z_1$
axis by (one of the branches of) the strong unstable manifold
$W^{\rm uu}(O)$; for $\eps\approx 0$ these two sets are very close to
one another since the slope of the eigenvector associated with the strong 
eigenvalue of the origin converges to $0$ as $\eps\to0$. Specifically, 
a solution is of type I if $(\eta, z_2(0))$
is above $W^{\rm uu}(0)$ and of type II if $(\eta, z_2(0))$ is below
$W^{\rm uu}(O)$. A solution of type I or II which passes near $N^+$ 
exists only if $(\eta, z_2(0))$ is above
$W^s(P_+)$. It follows that solutions of type I can be found as long
as $P_+$ is below $M_+$, while solutions of type II exist only if
$\mu>\mu_\ast$, i.e., $W^s(P_+)$ is below $W^{\rm uu}(O)$. This is
consistent with the discussion in the preceding paragraphs.

%

\begin{figure}
\centering
\includegraphics[width=\textwidth]{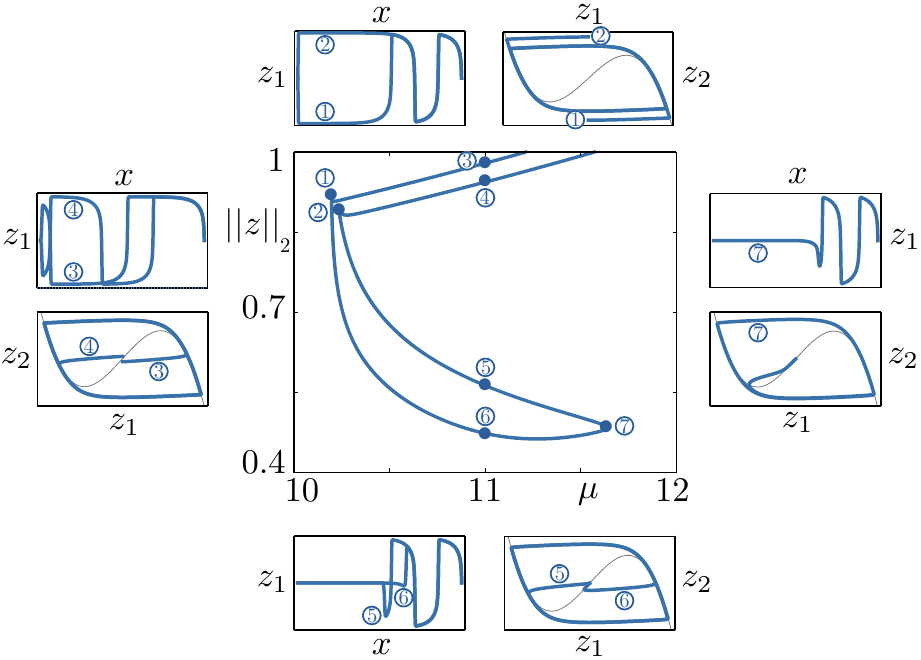}
\caption{Branch (central panel) corresponding to solutions that make a full turn
around $N$ before connecting back to $z_1=0$. The central panel shows a zoom of the
rightmost branch from Figure~\ref{mubrdetail}(a) corresponding to $\eta=-10^{-10}$.
Side panels show different solution profiles in space and in the phase plane at
different locations along this branch as indicated by the inscribed
integers.\label{double-turn}}
\end{figure}

In fact there is another class of solutions to the perturbed BVP, corresponding, in phase
space, to one or more full turns around the origin before returning to the $z_1$ axis
at $x=\ell$. These solutions follow the heteroclinic cycle mentioned above, and we can
understand their behavior using the same phase plane analysis. We focus on the branch
in the central panel of Figure~\ref{double-turn}. The properties of the solutions on
this branch can be classified by their behavior near the origin and near the saddle
points $P_{\pm}$. For example, the solution labeled 7 is at a fold point of the
branch. Like solution 5 from Figure~\ref{canard}, it contains a maximal canard
segment; the explanation of that fold point is similar to that provided earlier.
Solutions 5 and 6 in Figure~\ref{double-turn} occur after and before the transition
through the maximal canard, respectively. Solutions 1 and 2 are located at local
maxima on the branch and the associated $\mu$ values are both very close to the
(right) vertical segment on the branch shown in Figure~\ref{morecanards}(a); as in
the similar cases discussed earlier, these solutions come close to either of the two
saddle points, explaining the sharp increase in their $L_2$ norm. Lastly, solution 3
is close to a type II solution, and solution 4 is very similar to such a solution but
with $z_2(0)$ positive. Solutions with more turns are also possible and behave
analogously.

Finally, since the transition from the branch with $\eta=0$ to the branches with
$\eta\neq0$ can be seen as a broken pitchfork bifurcation, we expect to find
disconnected solution branches for $\mu>\mu_\mathrm{g}$ and $\eta\ne0$. These are indeed
present, as shown in Figure~\ref{brmularge} for $\eta=10^{-7}$ and $\eta=10^{-5}$.
The solution profiles on these branches correspond to orbit segments that make
several turns along $N$ before connecting back to $z_1=0$, and these lie, in the
limit $\eps\to0$, on additional solution branches of the unperturbed
BVP. In fact, for $\ell$ sufficiently large, there will be many such
solutions to the BVP considered by Chomaz \& Couairon, depending on the number of
turns they make along $N$ before connecting back to $z_1=0$. When the inlet
boundary condition becomes $z_1(0)=\eta\ne0$ with $\eta$ small (positive or
negative), one finds solution branches that follow the unperturbed ($\eta=0$)
branches corresponding to a given number of turns; such branches may also
connect two such unperturbed branches, as shown in Figure~\ref{brmularge}. Note
that all the branches that exist in the range $\mu>\mu_\mathrm{g}$ have regular limit
points without vertical segments, in contrast to the branches passing through
$\mu=\mu_{\ast}$ and $\mu=\mu_{c+}$. This is because for $\mu>\mu_\mathrm{g}$ the
unstable equilibrium point at the origin acquires complex eigenvalues (thereby
preventing orbit flips) so that the long-term dynamics of the ODE
system~\eqref{vdpAz1}--\eqref{vdpAz2} takes the form of an attracting limit cycle
surrounding the origin (thereby preventing any connection between the origin and the
  saddle points $P_{\pm}$). {Stability calculations show that these additional
  solutions are all unstable (in time).
\begin{figure}
\centering
\includegraphics{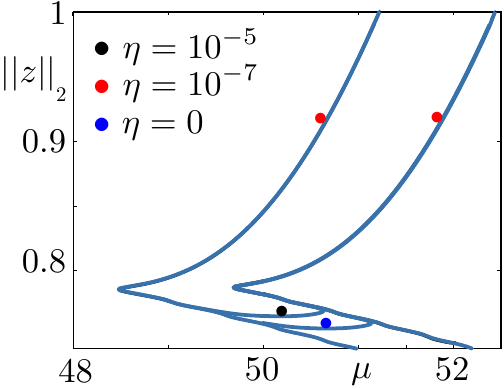}
\caption{Solution branches for $\mu>\mu_\mathrm{g}\approx 36.030$, for both $\eta=0$ and $\eta>0$
(see legend). The solutions resemble those shown in Figure~\ref{double-turn}, but
with more than one full turn along $N$. Unlike the type I and II branches described
earlier these branches do not exhibit any vertical segments. This is a consequence of
the fact that for $\mu>\mu_\mathrm{g}$ the origin $(0,0)$ has complex (spatial) eigenvalues.}
\label{brmularge}
\end{figure}

\section{Appendix: Numerical methods}\label{sec:numericalMethods}
This appendix summarizes the numerical technique used to obtain the results reported in this paper.
We discretize the PDE~\eqref{eq:PDEA} using a grid of evenly spaced points
$x = \{x_i\}_{i=1}^{n+1}$, with $x_i = i\ell/n$, and construct the approximation
vectors $A(t) = \{A_i(t)\}$, where $A_i(t) \approx A(x_i,t)$. We employ the
method of lines to obtain a large set of nonlinear ODEs
\begin{equation}\label{eq:discrPDE}
  \begin{aligned}
  A_1 & = \eta, \\
  \dot A_i & = \eps \mu A_i - \sqrt{\eps} A^3_i +
  \sum_{j=1}^{n+1} \Big[ (D^2_x)_{ij} - (1-\alpha A^2_i) (D_x)_{ij} \Big] A_j  \quad
  i = 2,\ldots,n, \\
  A_{n+1} & = 0,
  \end{aligned}
\end{equation}
where $D_x$ and $D^2_x$ are $(n-1) \times (n+1)$ finite-difference differentiation
matrices for the operators $\partial_x$ and $\partial_{xx}$, respectively. In all PDE
computations we use $n=4000$ gridpoints. Time stepping is performed using MATLAB's
\texttt{ode23s} with default parameters and explicitly formed Jacobian. For the
stochastic time stepping we use a standard Euler-Maruyama time-stepping scheme
\cite{Lord14} with $s=10^{-3}$. Steady states are computed using MATLAB's
\texttt{fsolve} with default parameter values and continued in parameter space using
the routines developed in~\cite{RankinAvitabile14}. Stability of a steady state
$A_*=\{A_*(x_i)\}$ is studied by computing the first $40$ eigenvalues of the problem
$M \psi = J(A_*) \psi$ with the smallest absolute value, where
\[
M = \mathrm{diag}(0,1,\ldots,1,0) \in \mathbb{R}^{(n+1)\times(n+1)},
\]
\[
J =
\begin{bmatrix}
  e_1 \\
  D^2_x - \mathrm{diag}(1 - \alpha A^2_*) D_x + \mathrm{diag}(2\alpha A_* D_x A_* + \eps \mu - 3 \sqrt{\eps} A^2_*) \\
  e_{n+1}
\end{bmatrix}
\in \mathbb{R}^{(n+1)\times(n+1)}
\]
and the $\{e_i\}$ are row vectors forming the canonical basis of $\mathbb{R}^{n+1}$.


\section*{Acknowledgments}
We thank C. Beaume, E. Hall and R. Thul for discussions. This work was supported in
part by the Engineering and Physical Sciences Research Council under Grant No.
EP/P510993/1 (DA) and the  National Science Foundation under Grants DMS-1211953 and
DMS-1613132 (EK).


\end{document}